\documentclass[twocolumn,final,aps,superscriptaddress,eqsecnum,10pt
amssymb,amsmath,amsfonts,showpacs,floatfix,citeautoscript]{revtex4}
\usepackage{graphicx,epstopdf}
\usepackage{subfigure}
\usepackage{mathrsfs}
\usepackage{bm}
\usepackage[latin1]{inputenc}
\usepackage{wasysym}
\usepackage{color}
\usepackage[colorlinks,bookmarks=false,citecolor=blue,linkcolor=red,urlcolor=blue]{hyperref}
\newcommand{\be}{\begin{equation}}
\newcommand{\ee}{\end{equation}}
\newcommand{\bea}{\begin{eqnarray}}
\newcommand{\eea}{\end{eqnarray}}
\newcommand{\dagga}{{\phantom{\dagger}}}
\newcommand{\vecr}{{\vec r}}

\newcommand{\veck}{{\vec k}}
\newcommand{\vecq}{{\vec q}}

\def\vec{\mathbf}
\begin{document}
\title{Semiclassical approach to ground-state properties of hard-core bosons in two dimensions}
\author{Tommaso Coletta}
\affiliation{Institute of Theoretical Physics, Ecole Polytechnique F\'ed\'erale de Lausanne (EPFL), CH-1015 Lausanne, Switzerland}
\author{Nicolas Laflorencie}
\affiliation{Laboratoire de Physique Th\'eorique, Universit\'e de Toulouse, UPS, (IRSAMC), F-31062 Toulouse, France}
\author{Fr\'ed\'eric Mila}
\affiliation{Institute of Theoretical Physics, Ecole Polytechnique F\'ed\'erale de Lausanne (EPFL), CH-1015 Lausanne, Switzerland}
\date{\today}


\begin{abstract}
Motivated by some inconsistencies in the way quantum fluctuations are included beyond the classical
treatment of hard-core bosons on a lattice in the recent literature, we revisit the large-$S$ semi-classical
approach to hard-core bosons on the square lattice at $T=0$. First of all, we show that, if one stays at
the purely harmonic
level, the only correct way to get the $1/S$ correction to the density is to extract it from the derivative
of the ground state energy with respect to the chemical potential, and that to extract it from a calculation
of the ground state expectation value of the particle number operator, it is necessary to include $1/\sqrt{S}$
corrections to the harmonic ground state. Building on this alternative approach to get $1/S$ corrections,
we provide the first semiclassical derivation of the momentum distribution, and we revisit the calculation
of the condensate density. The results of these as well as other physically relevant quantities such as
the superfluid density are systematically compared to quantum Monte Carlo simulations. This comparison shows
that the logarithmic corrections in the dilute Bose gas limit are only captured by the semi-classical approach if
the $1/S$ corrections are properly calculated, and that the semi-classical approach is able to reproduce
the $1/k$ divergence of the momentum distribution at $k=0$. Finally, the effect of $1/S^2$ corrections is briefly
discussed.
\end{abstract}

\maketitle

\section{Introduction}
Models of interacting bosons on a lattice are ubiquitous. They have been introduced
to describe the low energy physics of systems as different as thin superconducting films~\cite{Jaeger89},
Josephson junction arrays~\cite{Fazio01}, $^4$He on substrates~\cite{Matsubara56,Krauth91,Ceperley95}, cold atoms in optical lattices~\cite{Bloch08},
bipolarons~\cite{Alexandrov94}, or quantum magnets in a field~\cite{Affleck91,Giamarchi08}. In simple (unfrustrated) geometries
quantum Monte Carlo (QMC) simulations do not suffer from any minus sign problem, and the resulting picture is often quite
clear~\cite{Hebert02}. However, in many recent applications, the relevant effective model contains
terms that lead to a severe minus sign problem. This is for example true for the
bosonic description of frustrated quantum magnets in a magnetic field~\cite{Mila11}, where QMC approaches are not appropriate. To investigate
such models, it is important to develop alternative approaches.

A very important subclass is that of models of hard-core bosons on a lattice in
which the on-site repulsion is assumed to be infinite so that it is impossible to have
more than one boson at a given site. Such models appear for instance
very naturally in the description of dimer-based spin-1/2 quantum magnets in a field~\cite{Giamarchi08}.
In this paper, we will concentrate on a model of hard-core bosons on a two-dimensional
square lattice described by the simple Hamiltonian:
\begin{equation}\label{eq:Hardcore boson model}
\mathcal{H}=-t\sum_{\left\langle i,j\right\rangle}({a_i^\dagger a_j+a_i a_j^\dagger})-\mu\sum_i{n_i} ,
\end{equation}
where $t$ is the hopping amplitude between neighboring sites, $\mu$ is the chemical potential and $a_i^\dagger$ $(a_i)$ denotes the operator creating (destroying)
a hard-core boson at site $i$. One distinct property of hard-core boson models is that they can
be mapped exactly onto spin-1/2 models using the Matsuda-Matsubara transformation~\cite{Matsubara56}:
$n_i=S_i^z+1/2$, $a_i^\dagger=S_i^+$ and $a_i=S_i^-$.
The equivalent spin-1/2 model is a ferromagnetic $XY$ model with magnetic field $\mu$ pointing in the $z$ direction:
\begin{equation}\label{eq:Spin model}
\mathcal{H}=-t\sum_{\left\langle i,j\right\rangle}{2\left(S_i^x S_j^x + S_i^y S_j^y \right)}-\mu\sum_i\left(S_i^z+1/2\right)
\end{equation}
On the basis of this mapping, a semi-classical approximation can be developed starting from the large $S$ limit
of this spin Hamiltonian. This approach has been developed in a series of papers \cite{Scalettar95,Murthy97,Bernardet02}. The paper by Bernardet and coworkers~\cite{Bernardet02} includes a
careful comparison with QMC simulations
and shows that, already at the
order of linear-spin wave theory, the semi-classical approach is quantitatively accurate.

Building on this success, this semi-classical approach has recently been used quite systematically in
the investigation of frustrated models~\cite{Murthy97,Pich98,Laflorencie09,Schaffer09,Ueda10,Duric10,Laflorencie11} for which it is often the only available analytical approximation.
These studies have revealed a number of subtleties however in the implementation of the semi-classical
approximation. A recurrent problem concerns the calculation of the bosonic density as a function of the
chemical potential~\cite{Hen09,Hen10}, or equivalently of the magnetization as a function of the field. Bernardet {\it et al.}
have calculated the density as the opposite of the derivative of the energy with respect to the chemical
potential, which is equivalent to calculating the magnetization as the opposite of the derivative of the energy with
respect to the field. But one could in principle equally well calculate the magnetization as the expectation value
of the operator $S^z$ in the ground state. However, at the harmonic level, the two definitions do not
lead to the same answer, and it is not clear which definition should be preferred.
 In addition, in its current setting,
the semi-classical approach only allows to calculate in a systematic way quantities that can be derived
from the ground state energy, i.e. the density and the superfluid stiffness. For instance, no attempt has been made
so far to calculate other ground state properties such as the momentum distribution function.
Finally, no attempt to check the convergence of the $1/S$ expansion by calculating higher order corrections
has been made.

In the present paper, we address all these issues. First of all, we show that, to get the same result
using the two definitions of the density, one has to include $1/\sqrt{S}$ corrections to
the harmonic ground state. These corrections have to be included to get the correct answer to order
$1/S$ because they contribute at this order when calculating the expectation value of $S^z$. In the
low density limit, we also show that these corrections are crucial to get the logarithmic corrections
predicted long ago for interacting two-dimensional bosons. Building on this success, we use this corrected
ground state to calculate the momentum distribution function, and we show that it leads to a divergence
at zero momentum that agrees with QMC results. We also provide two complementary ways to calculate the
condensate at the order $1/S$: from the derivative of the energy with respect to a transverse field, and
by a calculation to the zero-momentum occupation factor using the perturbed ground-state. Finally,
we calculate the $1/S^2$ correction to the ground state energy and shows that it improves over the $1/S$
result, supporting the basic assumption of the semiclassical approach that the $1/S$ expansion is
well behaved even for $S=1/2$.

Let us emphasize that we agree with all the results of Ref.[\onlinecite{Bernardet02}] to order $1/S$. In that respect, the main objective of the
present paper is to show how these results can be obtained from perturbing the harmonic ground state, with
two new results: a clear answer regarding the appropriate way to calculate the expectation value of 
observables at the order $1/S$, and the first semiclassical calculation of the momentum distribution.

This paper is organized as follows.
In Sec.~\ref{sec:Linear SW} the model is treated in the context of linear spin wave theory.
Sec.~\ref{sec:Corrections Harmonic groundstate} is devoted to the semiclassical correction of the harmonic ground state,
and to the computation of several observables in this perturbed ground-state.
Sec.~\ref{sec:Comparison QMC results} presents a comparison of the spin wave results obtained with the results of QMC simulations.
Sec.~\ref{sec:Sum rule} discusses the validity in the context of a $1/S$ expansion of the sum rule which states that the total density
is equal to the sum of the condensate density and of the average momentum distribution function.
Sec.~\ref{sec:S2} presents some results obtained beyond the linear spin wave approximation.
A short conclusion is given in Sec.\ref{sec:Conclusion}. Finally, some details about the calculation
of the superfluid density and of the momentum distribution are given in Appendices A and B.

\section{Linear spin wave theory}\label{sec:Linear SW}

\subsection{The model}

To perform a semi-classical expansion, it will prove useful to extend the model of Eq.(\ref{eq:Spin model})
in two ways. First of all, we rescale the amplitudes in such a way that the various terms are of the same order
in the large $S$ limit while the Hamiltonian of Eq.(\ref{eq:Spin model}) is recovered for $S=1/2$. Secondly, and
more importantly, we introduce a transverse field $\Gamma\geq0$ in the $x$ direction. These modifications lead to the
Hamiltonian:
\begin{equation}\label{eq:Spin model studied}
\mathcal{H}=-\frac{t}{S^2}\sum_{\left\langle i,j\right\rangle}{\left(S_i^x S_j^x + S_i^y S_j^y \right)}-\frac{\mu}{S}\sum_i{S_i^z}-\frac{\Gamma}{S}\sum_i{S_i^x}
\end{equation}
The introduction of a transverse field $\Gamma$ turned out to be an essential ingredient in two
respects. On one hand, it allows one to calculate the condensate density as the opposite
of the derivative of the ground state energy with respect to $\Gamma$, hence to get an expression
that is correct to order $1/S$. On the other hand, it breaks the continuous $U(1)$ symmetry of the Hamiltonian of Eq.~(\ref{eq:Spin model}) and opens a gap in the spectrum of the model. Thanks to this gap, the correction to
the harmonic ground state is not divergent, and the corrected ground state can be used to calculate the
expectation value of various observables to order $1/S$. The results for the original model
are then obtained by taking the limit $\Gamma\rightarrow0$ of the expectation values.

\subsection{Classical solution}

In the classical limit, spin operators are replaced by three-dimensional vectors of norm $S$.
In the absence of a transverse field, the ground state consists of spins ordered ferromagnetically in the $x-y$ plane with a longitudinal magnetization $m$ that varies linearly with the magnetic field $\mu$ until saturation.
When $\Gamma > 0$, the classical solution lies in the $x-z$ plane and can be parametrized as follows:
\begin{equation}\label{eq:Classical spins}
\left(\begin{array}{c}
S_i^x\\
S_i^y\\
S_i^z\\
\end{array}\right)
=S
\left(\begin{array}{c}
\sin\theta\\
0\\
\cos\theta\\
\end{array}\right).
\end{equation}
With this parametrization, the classical energy per site is given by:
\begin{equation}\label{eq:Eclassical}
E^{(0)}=-2t\sin^2\theta-\mu\cos\theta-\Gamma\sin\theta.
\end{equation}
The angle $\theta$ is fixed by minimizing the classical energy,
\begin{equation}\label{eq:Condition on theta}
-4t\sin\theta\cos\theta+\mu\sin\theta-\Gamma\cos\theta=0.
\end{equation}
In the limit $\Gamma\rightarrow 0$, Eq.~(\ref{eq:Condition on theta}) has the simple solution
$\cos\theta_0=\mu/4t$,
and $S_i^x$ is different from zero in the field range $-4\leq \mu/t \leq 4$.
{This defines the critical magnetic field $\mu_c=-4t$ at which the system starts to acquire a transverse magnetization.
For $S=1/2$, at $\mu>\mu_c$, we have $S_i^z>-1/2$ which in terms of the original hardcore boson model corresponds to a non zero density of bosons.
Thus $\mu_c$ is the value of chemical potential at which the system is no longer empty and an hardcore boson population starts to develop.}
For $\Gamma\neq0$, the angle $\theta$ is a function of $\Gamma$.
In the following, we will focus on small $\Gamma$ case, and we will calculate the small $\Gamma$ correction
to several quantities. From the equation $\sin\theta(-4t\cos\theta+\mu)=\Gamma\cos\theta$, it is
easy to see that the first order correction to $\theta$ is given by:
\begin{equation}\label{eq:Theta to first order in gamma}
\left.\frac{\partial\theta}{\partial\Gamma}\right|_{\Gamma=0}=\frac{\cos\theta_0}{4t\sin^2\theta_0}.
\end{equation}
The classical on site magnetization is $S^z=S\cos\theta$ and the classical hardcore boson density is given by
$\rho^{\textrm{class.}}=(\cos\theta+1)/2$. In the limit where the transverse field vanishes the classical density is given by:
\begin{equation}\label{eq:Classical density}
\rho^\textrm{class.}=\frac{1}{2}\left(\frac{\mu}{4t}+1\right).
\end{equation}

\subsection{Holstein-Primakoff transformation}
In order to study the effect of quantum fluctuations around this classical solution, we start by performing a rotation of the spins at each site
\begin{equation}\label{eq:Local Spin Rotation}
 \begin{array}{lll}
  S_i^x &=& \cos\theta S_i^{x^\prime}+\sin\theta S_i^{z^\prime} \\
  S_i^y &=& S_i^{y^\prime} \\
  S_i^z &=& -\sin\theta S_i^{x^\prime}+\cos\theta S_i^{z^\prime}
 \end{array}
\end{equation}
such that the Hamiltonian of Eq.~(\ref{eq:Spin model studied}), expressed in the rotated frame $(x^\prime, y^\prime, z^\prime)$, has a ferromagnetic ground state.
The new spin operators can be expressed in terms of Holstein-Primakoff bosons~\cite{Holstein40}. To next
to leading order, the expressions take the form:
\begin{equation}\label{eq:HP transformation}
\begin{array}{lll}
S_i^{z^\prime}&=&\displaystyle S-b_i^\dagger b_i^\dagga \\[3mm]
S_i^{x^\prime}&=&\displaystyle \frac{\sqrt{2S}}{2} (b_i^\dagga+b_i^\dagger)-\frac{1}{4\sqrt{2S}}\left(n_ib_i^\dagga + b_i^\dagger n_i \right)+\ldots \\[3mm]
S_i^{y^\prime}&=&\displaystyle \frac{\sqrt{2S}}{2i}(b_i^\dagga-b_i^\dagger)-\frac{1}{4i\sqrt{2S}}\left(n_ib_i^\dagga - b_i^\dagger n_i \right)+\ldots
\end{array}
\end{equation}
The resulting Hamiltonian in terms of Holstein-Primakoff bosons can be expanded as
\begin{equation}\label{eq:HP expansion of H}
\mathcal{H}=\sum_{n\ge 0}\mathcal{H}^{(n)},
\end{equation}
where $\mathcal{H}^{(n)}$ is proportional to $S^{-\frac{n}{2}}$. The first term of this series is $\mathcal{H}^{(0)}=N E_{(0)}$,
$N$ being the total number of sites.
By construction, $\mathcal{H}^{(1)}=0$ since we expand around a spin configuration which is a classical minimum of the energy.
$\mathcal{H}^{(2)}$ is quadratic in bosonic operators while $\mathcal{H}^{(3)}$ and $\mathcal{H}^{(4)}$ contain only three or four boson terms respectively. Their expressions are given by:
\bea\label{eq:H2}
 \mathcal{H}^{(2)}= &-&\frac{t}{2S}\sum_{\langle i,j\rangle}{\left(\cos^2\theta+1\right)\left(b_i^\dagga b_j^\dagger+b_i^\dagger b_j^\dagga\right)}\nonumber\\
                  &-&\frac{t}{2S}\sum_{\langle i,j\rangle}{\left(\cos^2\theta-1\right)\left(b_i^\dagga b_j^\dagga+b_i^\dagger b_j^\dagger\right)}\\
                  &+&\frac{1}{S}\sum_i {b_i^\dagger b_i^\dagga\left( 4t\sin^2\theta + \mu\cos\theta +\Gamma\sin\theta\right)}, \nonumber
\eea
\begin{equation}\label{eq:H3}
 \mathcal{H}^{(3)}=\displaystyle \frac{2t}{S\sqrt{2S}}\sum_{\langle i,j\rangle}{n_i(b_j^\dagga+b_j^\dagger) \sin\theta\cos\theta},
\end{equation}
\bea
\label{eq:H4}
 \mathcal{H}^{(4)}=&-&\frac{t}{S^2}\sum_{\langle i,j\rangle}{\frac{1}{8}(1-\cos^2\theta)\left([n_i+n_j]b_ib_j+\textrm{h.c.}\right)} \nonumber\\
&-&\frac{t}{S^2}\sum_{\langle i,j\rangle}{-\frac{1}{8}(1+\cos^2\theta)\left(b_i^\dagger[n_i+n_j]b_j^\dagga+\textrm{h.c.}\right)}\nonumber\\
                  & -&\frac{t}{S^2}\sum_{\langle i,j\rangle}{\sin^2\theta~n_in_j}.
\eea
The calculation of $1/S$ corrections, to which most of the paper is devoted, is based on $\mathcal{H}^{(2)}$
and $\mathcal{H}^{(3)}$. The fourth order correction $\mathcal{H}^{(4)}$ will only be used in Section~\ref{sec:S2}
when we calculate the $1/S^2$ correction to the energy.
\subsection{Diagonalization of the harmonic Hamiltonian}
In terms of the Fourier transformations of the Holstein-Primakoff operators defined by
\begin{equation}\label{eq:Definition FT}
 \begin{array}{lll}
  b_j=\displaystyle \frac{1}{\sqrt{N}}\sum_{\veck}{b_\veck e^{i \vec{R}_j\vec{k}}} & \quad
  b_j^\dagger=\displaystyle \frac{1}{\sqrt{N}}\sum_\veck{b_\veck^\dagger e^{-i \vec{R}_j\vec{k}}} \\[3mm]
  b_\veck=\displaystyle \frac{1}{\sqrt{N}}\sum_j{b_j e^{-i \vec{R}_j\vec{k}}} & \quad
  b_\veck^\dagger=\displaystyle \frac{1}{\sqrt{N}}\sum_j{b_j^\dagger e^{i \vec{R}_j\vec{k}}}
 \end{array}
\end{equation}
$\mathcal{H}^{(2)}$ can be decoupled into a sum over different modes,
\begin{equation}\label{eq:H2 FT}
\begin{array}{lll}
 \mathcal{H}^{(2)}&=&\displaystyle \frac{1}{S} \sum_{\veck}{(b_\veck^\dagger, b_{-\veck}^\dagga)
                     \left(\begin{array}{cc}
                         A_\veck & B_\veck \\
                         B_\veck & A_\veck
                         \end{array}
                     \right)
                     \left(\begin{array}{c}
                         b_\veck^\dagga\\
                         b_{-\veck}^\dagger
                         \end{array}
                     \right)}  \\
                  &-&\displaystyle \frac{1}{2S}\sum_{\veck}(4t\sin^2\theta+\mu\cos\theta+\Gamma\sin\theta)
\end{array}
\end{equation}
where the coefficients $A_\veck$ and $B_\veck$ are defined by:
\bea\label{eq:Coefficients A and B}
  A_\veck&=&\displaystyle -\frac{t}{2}\gamma_\veck(\cos^2\theta+1)+2t\sin^2\theta+\frac{\mu}{2}\cos\theta+\frac{\Gamma}{2}\sin\theta\nonumber\\
  B_\veck&=&\displaystyle \frac{t}{2}\gamma_\veck\sin^2\theta,
\eea
with $\gamma_\veck=\cos k_x+\cos k_y$. With the help of Eq.(\ref{eq:Theta to first order in gamma}), these
coefficients can easily be expanded to linear order in $\Gamma$:
\begin{equation}
 \begin{array}{lll}
  A_\veck&\approx&\displaystyle A^0_\veck+\frac{\Gamma}{2}\left[\frac{\cos^2\theta_0}{\sin\theta_0}\frac{\gamma_\veck}{2}+\frac{1}{\sin\theta_0}\right]\\
\\
  B_\veck&\approx&\displaystyle B^0_\veck+\Gamma\frac{\cos^2\theta_0}{4\sin\theta_0}\gamma_\veck
 \end{array}
\end{equation}
where $A^0_\veck=-t\left[\gamma_\veck\left(1+\cos^2\theta_0\right)-4\right]/2$ and
$B^0_\veck=\left[t\gamma_\veck\sin^2\theta_0\right]/2$ denote the coefficients $A_\veck$ and $B_\veck$ in the absence of a transverse field~\cite{Bernardet02}.
The second term in Eq.~(\ref{eq:H2 FT}) can also be expanded to first order in $\Gamma$, leading to $-(1/S)\left[\sum_{\veck}{2t+\Gamma/(2\sin\theta_0)}\right]$.

The quadratic Hamiltonian (\ref{eq:H2 FT}) can be diagonalized via a Bogoliubov transformation:
\begin{equation}
\begin{array}{ccc}
 b_\veck=u_\veck\alpha_\veck-v_\veck\alpha_{-\veck}^\dagger & & b_\veck^\dagger=u_\veck\alpha_\veck^\dagger-v_\veck\alpha_{-\veck}.
\end{array}
\end{equation}
The coefficients which ensure that the operators $\alpha_\veck^\dagga(\alpha_\veck^\dagger)$ satisfy bosonic commutation relations and that the Hamiltonian is diagonal are given by:
\begin{equation}
 \begin{array}{c}
  \displaystyle u_\veck^2=\frac{1}{2}\left(\frac{A_\veck}{\sqrt{A_\veck^2-B_\veck^2}}+1\right)\\[4mm]
  \displaystyle v_\veck^2=\frac{1}{2}\left(\frac{A_\veck}{\sqrt{A_\veck^2-B_\veck^2}}-1\right).
 \end{array}
\end{equation}
In terms of the Bogoliubov operators, and to first order in $\Gamma$, the Hamiltonian takes the diagonal form:
\begin{equation}\label{eq:H2 BT}
\begin{array}{lll}
 \mathcal{H}^{(2)}&=&\displaystyle
 \frac{2}{S}\sum_{\veck}{\sqrt{A_\veck^2-B_\veck^2}\ \alpha_\veck^\dagger\alpha_\veck^\dagga}\\
                  &+&\displaystyle\frac{1}{S}\sum_{\veck}{\left[\sqrt{A_\veck^2-B_\veck^2}-2t-\frac{\Gamma}{2\sin\theta_0}\right]}.
\end{array}
\end{equation}
The ground state of the harmonic Hamiltonian $\mathcal{H}^{(2)}$ is the vacuum of $\alpha$ particles.
We will refer to it as the harmonic ground state in the rest of this paper.
The first order $1/S$ correction to the energy per site is given by:
\begin{equation}\label{eq:Corrected energy to first order}
 E^{(2)}=\frac{1}{SN}\sum_{\veck}{\left[\sqrt{A_\veck^2-B_\veck^2}-2t-\frac{\Gamma}{2\sin\theta_0}\right]}.
\end{equation}
The only difference with the approach of Ref.~\onlinecite{Bernardet02} is that, as long as the
transverse field is strictly positive, the Bogoliubov transformation is well
behaved even at $\vec{k}=0$ since the excitation spectrum is gapped. Indeed,
for small $\veck$ and $\Gamma$, the excitation energy $\Omega_\veck=2\sqrt{A_\veck^2-B_\veck^2}/S$
can be written as
\be
\Omega_\veck\approx\sqrt{\Delta^2+v^2\veck^2},
\ee
with
\be
\Delta=\frac{2}{S}\sqrt{\Gamma t\sin\theta_0}+\mathcal{O}(\Gamma^\frac{3}{2}),
\end{equation}
and
\be
v=\frac{2}{S}t\sin\theta_0+\frac{\Gamma(1+3\cos^2\theta_0)}{4S\sin^2\theta_0}+\mathcal{O}(\Gamma^2).
\ee
In the limit $\Gamma\to 0$, the spectrum becomes gapless and linear, as expected for phonon-like excitations in a superfluid.
%
%
\subsection{Calculation of the densities from the ground state energy}

A system of bosons is characterized by three densities: the total density, the condensate density,
and the superfluid density. They can all be calculated as derivatives of the ground state energy.
Using the Hellman-Feynman theorem which states that
\begin{equation}
\left\langle \frac{\partial \mathcal{H}(h)}{\partial h}\right\rangle = \frac{\partial}{\partial h} {\left\langle \mathcal{H}(h)\right\rangle}
\end{equation}
where $h$ is some parameter of the Hamiltonian, one can calculate the longitudinal magnetization $m$
as
\be
m(S)=-S\frac{\partial E^{(2)}(\Gamma=0)}{\partial \mu}
\ee
and the transverse magnetization $m_\perp$ as
\be
m_\perp(S)=-S \left.\frac{\partial E^{(2)}}{\partial \Gamma}\right|_{\Gamma=0}
\ee
while the spin stiffness is given by the second derivative of the energy with respect to a twist
(see Appendix A).
The advantage of deriving these densities from the ground state energy is that, once we have an expression
of the energy to a given order in $1/S$, we obtain expressions of the densities which are correct at the
same order. Let us discuss the result for the various densities.

\subsubsection{Total density}

Using the expression of the energy of Eq.(\ref{eq:Corrected energy to first order}) for $\Gamma=0$, the derivative with respect to $\mu$ leads to the longitudinal magnetization
\be
\label{eq:Corrected Sz}
 m(S)= S\cos\theta_0+\frac{\cos\theta_0}{4}\frac{1}{N}\sum_{\veck}{\gamma_\veck\sqrt{\frac{{A_\veck^0-B_\veck^0}}{{A_\veck^0+B_\veck^0}}}}\nonumber
\ee
The total density $\rho$ is related to the longitudinal magnetization by $\rho = m(S=1/2)+1/2$, which leads to
\be
\label{eq:CorrectedDensity}
\rho=\frac{1+\cos\theta_0}{2}+\frac{\cos\theta_0}{4}\frac{1}{N}\sum_{\veck}{\gamma_\veck\sqrt{\frac{{A_\veck^0-B_\veck^0}}{{A_\veck^0+B_\veck^0}}}}
\ee
in perfect agreement with Ref.~\onlinecite{Bernardet02}.

\subsubsection{Condensate density}

Taking now the derivative of the energy of Eq.(\ref{eq:Corrected energy to first order}) with respect to $\Gamma$,
we obtain the following expression for the transverse magnetization:
\bea
 m_\perp(S) &=&  S\sin\theta_0\nonumber\\
 &-&\frac{1}{2\sin\theta_0}\frac{1}{N}\sum_\veck{\left(\frac{A_\veck^0}{\sqrt{(A_\veck^0)^2-(B_\veck^0)^2}}-1\right)}\nonumber\\
 &-&\frac{\cos^2\theta_0}{4\sin\theta_0}\frac{1}{N}\sum_{\veck}{\gamma_\veck\sqrt{\frac{{A_\veck^0-B_\veck^0}}{{A_\veck^0+B_\veck^0}}}}\nonumber
\eea
The condensate density $\rho_0$, which is the number of bosons occupying the $\vec{k}=0$ mode per site $\rho_0= \langle a_{\veck=0}^\dagger a_{\veck=0}^\dagga\rangle /N
       =\sum_{ij}{\left\langle S_i^+S_j^-\right\rangle}/N^2$, is simply
related to the transverse magnetization by $\rho_0=[m_\perp(S=1/2)]^2$, which leads to the expression
\bea
\label{eq:Corrected Condensate}
\rho_0&=& \frac{1}{4}\sin^2\theta_0-\frac{1}{2N}\sum_\veck{\left(\frac{A_\veck^0}{\sqrt{(A_\veck^0)^2-(B_\veck^0)^2}}-1\right)}\nonumber\\
       &-&\frac{\cos^2\theta_0}{4}\frac{1}{N}\sum_{\veck}{\gamma_\veck\sqrt{\frac{{A_\veck^0-B_\veck^0}}{{A_\veck^0+B_\veck^0}}}}.
\eea
This expression is different from that of Ref.~\onlinecite{Bernardet02}. The two expressions are strictly equivalent
only at $\mu=-4t$ (low density limit) and at $\mu=0$ (half filling).
In the range $-4<\mu/t<0$ and $0<\mu/t<4$ the expression of Ref.~\onlinecite{Bernardet02} differs form that of Eq.~(\ref{eq:Corrected Condensate})
by a term which is of order $1/S^2$ \footnote{Note that this difference comes from the fact that in Ref.~\cite{Bernardet02} both the condensate and superfluid
densities are expressed in terms of $\rho(1-\rho)$ where $\rho$ denotes the particle density corrected to order $1/S$ (\textit{G. Batrouni, private communication}).}.
We believe that the above expression for the transverse magnetization
is the correct one to order $1/S$. This will be further supported by a direct calculation of the expectation
value of $S^x$ in the next section.

\subsubsection{Superfluid density}

As explained in the appendix~\ref{sec:app}, the superfluid density is given at $1/S$ order by the following expression
\bea
\rho_{\rm sf}&=&\frac{1}{4}\sin^2\theta_0+\frac{1}{4Nt}\sum_{\veck}\left(2t-\sqrt{(A_{\veck}^0)^2-({{B}}_{\veck}^0)^2}\right)\nonumber\\
&-&\frac{\cos^2\theta_0}{4}\frac{1}{N}\sum_{\veck}{\gamma_\veck\sqrt{\frac{{A_\veck^0-B_\veck^0}}{{A_\veck^0+B_\veck^0}}}}.
\label{eq:RHOSFSW}
\eea
This expression is equivalent to order $1/S$ to the expression of Ref.~\onlinecite{Bernardet02}.
Looking at the expressions Eqs.~\eqref{eq:Corrected Condensate} and \eqref{eq:RHOSFSW} for the condensed and superfluid densities,
one can make several interesting observations. First, quantum fluctuations deplete the condensate whereas they enhance superfluidity.
In Eq.~\eqref{eq:RHOSFSW}, two contributions in the mechanism of superfluidity enhancement are present:
the first term comes from the nearest neighbor kinetic energy which increases (in absolute value) due to quantum fluctuations (see Eq.~\eqref{eq:Corrected energy to first order}),
and the second one is due to the increase of the total density of particle $\rho$ as seen in Eq.~\eqref{eq:CorrectedDensity}.
Interestingly, the same term appears in the condensate density, but with an opposite sign.

\section{Large $S$ corrections to the harmonic ground state}\label{sec:Corrections Harmonic groundstate}

Now that we have expressions for the longitudinal and transverse magnetizations valid to order $1/S$, let us show
how these expressions can be obtained as expectation values of $S^z$ and $S^x$. Since the expressions to order
$1/S$ have been derived from the energy calculated at the harmonic level, one might expect that it is sufficient
to calculate the expectation value of $S^z$ and $S^x$ in the harmonic ground state. As we shall see, this is not
the case. The basic reason is quite simple: in terms of Holstein-Primakoff bosons, the operators $S^z$ and $S^x$
contains terms of order $O(S)$ and terms of order $O(1)$. To get an expression which is correct up to order $O(1)$,
i.e. which includes all corrections up to $1/S$, one should thus include in the ground state corrections up to
order $1/\sqrt{S}$, if any, since the expectation value of the $O(S)$ part of the operators in such a correction will
give a contribution of order $O(1)$. As we shall now show, the term $\mathcal{H}^{(3)}$ in the expansion of the
Hamiltonian indeed leads to a correction to the ground state of order $1/\sqrt{S}$.

\subsection{Beyond the harmonic ground state}
The objective of this part is to compute the large $S$ corrections to the harmonic ground state.
To do so, we treat $\mathcal{H}^{(3)}$ Eq.~\eqref{eq:H3} as a perturbation to $\mathcal{H}^{(2)}$, the small
parameter being $1/S$.
The ground state of $\mathcal{H}^{(2)}$, the vacuum of $\alpha$ quasiparticles, being non degenerate, we use
Rayleigh-Schr\"odinger non degenerate perturbation theory.
To first order in perturbation, the perturbed ground state is given by:

\begin{equation}\label{eq:Perturbation theory}
\begin{array}{lll}
 |\psi\rangle &=& \displaystyle |0\rangle+\sum_{|e\rangle}\underbrace{\frac{1}{E_{|0\rangle}-E_{|e\rangle}}}_{\mathcal{O}(S)}\underbrace{{\langle e|\mathcal{H}^{(3)}|0\rangle|e\rangle}}_{\mathcal{O}(\frac{1}{S\sqrt{S}})}\\
              &=& \displaystyle |0\rangle+\frac{1}{\sqrt{S}}|\phi^{\frac{1}{2}}\rangle + \mathcal{O}(\frac{1}{S}),
\end{array}
\end{equation}
where $|0\rangle$ denotes the vacuum of $\alpha$ quasiparticles, $|e\rangle$ magnon excitations and $E_{|0\rangle}(E_{|e\rangle})$ the energy of the vacuum (excited states). The first correction to the
ground state is of order $1/\sqrt{S}$ since $1/(E_{|0\rangle}-E_{|e\rangle})$ is of order $\mathcal{O}(S)$ while
$\langle e|\mathcal{H}^{(3)}|0\rangle|e\rangle$ is of order $\mathcal{O}(1/S^{\frac{3}{2}})$. The second line defines $|\phi^{\frac{1}{2}}\rangle$, the ket that gives the $1/\sqrt{S}$ correction to the ground state.
The above wave function is correct to order $1/\sqrt{S}$ since all terms $\mathcal{H}^{(n)}$ with $n\geq4$ in the Holstein-Primakoff expansion (\ref{eq:HP expansion of H})
will contribute corrections of higher order in $1/S$.
Finally, we have to normalize the state, which leads to:
\begin{equation}\label{eq:Corrected wavefunction}
 \left| \psi_0 \right\rangle\approx\left(1-\frac{C}{2S}\right)\left| 0 \right\rangle+\frac{1}{\sqrt{S}}| \phi^{\frac{1}{2}} \rangle +\dots,
\end{equation}
where $C=\langle\phi^{\frac{1}{2}}|\phi^{\frac{1}{2}}\rangle$ and $\langle\psi_0|\psi_0\rangle=1+\mathcal{O}(1/S^2)$.
To compute $|\phi^{\frac{1}{2}}\rangle$, we first express $\mathcal{H}^{(3)}$ in Fourier space
\begin{equation}\label{eq:Fourier Trasform of H3}
 \mathcal{H}^{(3)}=\frac{2t}{S\sqrt{2NS}}\sum_{\veck,\vecq}{\sin\theta\cos\theta\gamma_{\vecq}\left(b_{\vecq+\veck}^\dagger b_\veck^\dagga b_\vecq^\dagga+b_{\veck-\vecq}^\dagger b_\veck^\dagga b_\vecq^\dagger\right)}
\end{equation}
with
\bea
\label{eq:Excitations H3}
  b_{\vecq+\veck}^\dagger b_\veck^\dagga b^\dagga_\vecq &=&
  (u_{\veck+\vecq}\alpha_{\veck+\vecq}^\dagger-v_{\veck+\vecq}\alpha_{-\veck-\vecq}^\dagga)\\
  &\times&(u_\veck\alpha_\veck^\dagga-v_\veck\alpha_{-\veck}^\dagger)(u_\vecq\alpha_\vecq^\dagga-v_\vecq\alpha_{-\vecq}^\dagger) \nonumber\\
\nonumber\\
 b_{\veck-\vecq}^\dagger b_\veck^\dagga b_\vecq^\dagger &=&
  (u_{\veck-\vecq}\alpha_{\veck-\vecq}^\dagger-v_{\veck-\vecq}\alpha_{-\veck+\vecq}^\dagga)\\
  &\times&(u_\veck\alpha_\veck^\dagga-v_\veck\alpha_{-\veck}^\dagger)(u_\vecq\alpha_\vecq^{\dagger}-v_\vecq\alpha_{-\vecq}^\dagga) \nonumber.
\eea
$\mathcal{H}^{(3)}$ being a three-body operator, its effect on the vacuum is to create one-magnon or three-magnon excitations.
In Sec.~\ref{sec:Averages in the corrected gs}, we will show that, for the computation of first order corrections to the average values
of the observables of interest in this paper, only single magnon excitations are relevant.
We thus write $| \phi^{\frac{1}{2}} \rangle$ as a sum of one-magnon and three-magnon contributions:
\be
| \phi^{\frac{1}{2}} \rangle = | \phi^{\frac{1}{2}} \rangle_{1m} + | \phi^{\frac{1}{2}} \rangle_{3m}
\ee
and we concentrate on the expression of the one-magnon contribution $| \phi^{\frac{1}{2}} \rangle_{1m}$.
Due to momentum conservation in Eq.~(\ref{eq:Fourier Trasform of H3}), the only single particle excitations
allowed have zero momenta.
Hence, $E_{|0\rangle}-E_{|e\rangle}=-\Omega_0=-2\sqrt{A_0^2-B_0^2}/S$ and $|\phi^\frac{1}{2}\rangle_{1m}$
is given by:
\begin{equation}
\begin{array}{ccl}
|\phi^{\frac{1}{2}}\rangle_{1m}&=&\displaystyle -\frac{t}{\sqrt{2}}\sin\theta\cos\theta\frac{(u_0-v_0)}{\sqrt{A_0^2-B_0^2}} \\
                          & &\displaystyle \quad \cdot \frac{1}{\sqrt{N}}\sum_{\veck}{\left[2v_\veck^2+\gamma_\veck(v_\veck^2-v_\veck u_\veck)\right]}|1_{\vecq=0}\rangle
\end{array}
\end{equation}
where $|1_{\vecq=0}\rangle$ denotes an excited state of one magnon with momenta $\vec{q}=0$.
Note that it is only possible to write down such an expression because we have included a transverse field
in the Hamiltonian, so that the Bogoliubov transformation is not singular at $\vec{k}=0$. This
expression actually diverges in the limit $\Gamma \rightarrow 0$ because $\sqrt{A_0^2-B_0^2}=O(\Gamma^{1/2})$
while  $(u_0-v_0)=O(\Gamma^{1/4})$. As we shall see, the limit $\Gamma \rightarrow 0$ must be taken
after calculating the expectation value of the operators.

\subsection{Expectation values of observables}
\label{sec:Averages in the corrected gs}
\subsubsection{Total density}
Let us first use the perturbed ground state to calculate the expectation value of $S^z$.
The first step is to express $\langle S^z\rangle$ in terms of the spin operators in the rotated frame, and
to use the expansion of these operators in terms of Holstein-Primakoff bosons. This leads to:
\begin{equation}\label{eq:Average Sz}
 \begin{array}{lll}
  \langle S^z_i\rangle & = & \cos\theta \langle S^{z^\prime}_i\rangle - \sin\theta \langle S_i^{x^\prime}\rangle  \\[3mm]
                     & = & \cos\theta \left(S-\langle b_i^\dagger b_i \rangle\right) \\[1mm]
                     &   & - \sin\theta \langle\frac{\sqrt{2S}}{2}(b_i+b_i^\dagger)-\frac{1}{4\sqrt{2S}}(n_ib_i+b_i^\dagger n_i)\rangle
 \end{array}
\end{equation}
At the classical level, the average magnetization is given by $S\cos\theta_0$ when $\Gamma=0$.
The spin wave corrections to this result are of order $\mathcal{O}(1)$. Thus, given the structure of the perturbed ground-state
$|\psi_0\rangle\approx\left(1-C/(2S)\right)\left| 0 \right\rangle+S^{-\frac{1}{2}}| \phi^{\frac{1}{2}} \rangle$,
the terms entering the average magnetization to order $\mathcal{O}(1)$ are:
\begin{equation}\label{eq:Relevant terms Sz}
\begin{array}{lll}
 \langle S_i^z\rangle|_{\Gamma=0}&=&\displaystyle S\cos\theta_0-\lim_{\Gamma\rightarrow0} \left\{\cos\theta\langle0|b_i^\dagger b_i|0\rangle \right. \\
                   & &\quad      \left. +\sin\theta \frac{1}{\sqrt{2}} \left(\langle 0|b_i^\dagger+b_i |\phi^{\frac{1}{2}}\rangle_{1m}+\textrm{h.c.}\right)\right\}.
\end{array}
\end{equation}
In the above expression, we have only included $|\phi^{\frac{1}{2}}\rangle_{1m}$ in the matrix element
of $b_i^\dagger+b_i$ since the operator $b_i^\dagger+b_i $ can at most create or destroy one Bogoliubov excitation.
The three magnon component $|\phi^\frac{1}{2}\rangle_{3m}$ would only contribute to the matrix element $(1/\sqrt{S})\langle0|(n_ib_i+b_i^\dagger n_i)/(4\sqrt{2S})|\phi^\frac{1}{2}\rangle$, but this term is of
order $\mathcal{O}(1/S)$ and can be neglected since we are interested in the $O(1)$ correction to the
expectation value of $S^z$.

The matrix elements $\langle 0|b_i^\dagger b_i^\dagga |0 \rangle$ and $\langle 0|b_i^\dagger+b_i^\dagga |\phi^{\frac{1}{2}}\rangle_{1m}$ are readily computed in Fourier space:
\begin{equation}\label{eq:Matrix element 1}
\langle 0|b_i^\dagger b_i^\dagga |0 \rangle=\frac{1}{N}\sum_{\veck}{v_{\veck}^2},
\end{equation}
and
\begin{equation}\label{eq:Matrix element 2}
\begin{array}{c}
 \langle 0|b_i^\dagger+b_i^\dagga |\phi^{\frac{1}{2}}\rangle_{1m}=\displaystyle \frac{1}{\sqrt{N}} \sum_{\veck^\prime} \langle 0|(b_{\veck^\prime}^\dagger e^{-i\veck^\prime {\vecr}_i}+b_{\veck^\prime}^\dagga e^{i\veck^\prime \vecr_i})|\phi^{\frac{1}{2}}\rangle_{1m}  \\[4mm]
                                                      = \displaystyle \frac{1}{\sqrt{N}} \langle 0|\left(u_0-v_0\right)\alpha_0|\phi^{\frac{1}{2}}\rangle_{1m} \\ [4mm]
 = \displaystyle -\frac{t}{\sqrt{2}}\sin\theta\cos\theta\frac{1}{N}\sum_{\veck\neq0}{\frac{2v_\veck^2+\gamma_\veck(v_\veck^2-v_\veck u_\veck)}{A_0+B_0} }
\end{array}
\end{equation}
with
\be
A_0+B_0=2t\sin^2\theta_0+\Gamma\left(\frac{\cos^2\theta_0}{\sin\theta_0}
+\frac{1}{2\sin\theta_0}\right)+\mathcal{O}(\Gamma^2).
\ee
Note that the expression of the second matrix element has a finite $\Gamma \rightarrow 0$ limit because of the
extra $u_0-v_0$ factor.
Injecting back Eqs.~(\ref{eq:Matrix element 1}) and (\ref{eq:Matrix element 2}) into Eqs.~(\ref{eq:Relevant terms Sz}), one recovers exactly the expression of the longitudinal magnetization obtained before from the derivative of the
energy calculated at the harmonic level.

\subsubsection{Condensate density}

The same procedure can be repeated for the operator $\langle S_i^x\rangle=\sin\theta \langle S_i^{z^\prime}\rangle + \cos\theta \langle S_i^{x^\prime}\rangle$, and the $\Gamma \rightarrow 0$ limit of its expectation value
is given by:
\begin{equation}\label{eq:Relevant terms Sx}
\begin{array}{lll}
 \langle S_i^x\rangle|_{\Gamma=0}&=&\displaystyle S\sin\theta_0 -\lim_{\Gamma\rightarrow 0} \left\{\sin\theta\langle0|b_i^\dagger b_i^\dagga|0\rangle \right. \\
                   & &\quad      \left. -\cos\theta \frac{1}{\sqrt{2}} \left(\langle 0|b_i^\dagger+b_i^\dagga |\phi^{\frac{1}{2}}\rangle_{1m}+\textrm{h.c.}\right)\right\}.
\end{array}
\end{equation}
Injecting back Eqs.~(\ref{eq:Matrix element 1}) and (\ref{eq:Matrix element 2})
into this expression leads to exactly the same expression for the condensate
as the one obtained from the derivative of the energy with respect to $\Gamma$.

\subsubsection{Momentum distribution}

The main advantage of this approach is that it gives access to observables that cannot be calculated
as derivatives of the energy. Among them, a physically very important one is the momentum distribution
defined by:
\begin{equation}\label{eq:distribution n(k)}
 \langle a_{\veck}^\dagger a_{\veck}\rangle=\lim_{\substack{\Gamma\rightarrow0 \\ S \rightarrow \frac{1}{2}}}\frac{1}{N}\sum_{ij}{\left\langle S_i^+S_j^-\right\rangle e^{i\veck(\vecr_i-\vecr_j)}}.
\end{equation}
The details of the calculation of $\langle S_i^+S_j^-\rangle$ in the perturbed ground state (\ref{eq:Corrected wavefunction}) are given in Appendix~\ref{sec:app2}. For $\vec{\veck}\neq0$, the momentum distribution is given by:
\bea
\label{eq:n1(k)}
\langle a_\veck^\dagger a_\veck^\dagga\rangle&=& \frac{1}{4}(1+\cos\theta_0)^2\nonumber\\
&+&\frac{1}{2}\left[(1+\cos^2\theta_0)v_\veck^2+u_\veck v_\veck\sin^2\theta_0\right].
\eea
The classical expression for $\langle a_\veck^\dagger a_\veck^\dagga\rangle$ is  equal to the square of the classical density and does not depend on $\veck$.
Eq.~\eqref{eq:n1(k)} can be re-expressed as
\bea
\langle a_\veck^\dagger a_\veck^\dagga\rangle=\frac{2t\left[1+\cos^2\theta_0(1-\gamma_\veck)\right]}{\Omega_\veck}+\frac{\cos \theta_0}{2},
\label{eq:NKSW}
\eea
from which one sees that the $1/S$-corrected distribution diverges like $1/\Omega_{\veck}$ when approaching the condensate point at $\vec{k}=0$.
More precisely, the momentum distribution is singular and behaves like $\sim \sin\theta_0/k$ for all values of the field $-4<\mu/t<4$.
The integral of this quantity over the whole Brillouin zone is convergent and yields the number of uncondensed particles, which is given by
\begin{equation}\label{eq:Corrected density of uncondensed particles}
\begin{array}{lll}
\displaystyle\frac{1}{N}\sum_{\veck\neq0}\langle a_\veck^\dagger a_\veck^\dagga\rangle  &=& \displaystyle \left(\rho^{\rm class.}\right)^2\left(1-\frac{1}{N}\right) \\
                                                        & &\displaystyle+\frac{1}{N}\sum_{\veck\neq0} \frac{1}{2}\cos^2\theta_0(v_\veck^2-u_\veck v_\veck)\\
                                                        & &\displaystyle+\frac{1}{N}\sum_{\veck\neq0} \frac{1}{2}(v_\veck^2+u_\veck v_\veck)\\
\end{array}
\end{equation}

Let us point out that one can also recover the condensate density using Eq.~(\ref{eq:FT distribution n(k)}) in Appendix~\ref{sec:app2}. Indeed,
if $\vec{k}$ is replaced by zero in the right hand side of Eq.~(\ref{eq:FT distribution n(k)}),
the expression can be rearranged to lead again to the condensate density obtained previously using the definition $\rho_0=\langle S^x\rangle^2$.

\section{Comparison with QMC simulations}\label{sec:Comparison QMC results}
In this section we compare the large $S$ approximation of various quantities to exact QMC estimates obtained using the Stochastic Series Expansion (SSE) algorithm~\cite{Sandvik02}.
The simulations have been performed for the hard-core boson model Eq.~\eqref{eq:Hardcore boson model} on square lattices $L \times L$ with $L=8,~16,~24,~32$ at temperatures low enough to get ground state estimates ($\beta/t\propto L^2$),
in particular in the dilute limit where the finite size gap scales as $\sim L^{-z}$ with $z=2$.
Overall, we agree with the numerical results of Ref.[\onlinecite{Bernardet02}] whenever
we could compare. We have nevertheless included numerical results for the density, the condensate 
and the superfluid density to be able to discuss their behaviour close to $\mu_c$, where the corrections
to the harmonic ground state turn out to be crucial. In addition to these quantities that had already
been discussed in Ref.[\onlinecite{Bernardet02}], this section also contains QMC results for the
momentum distribution.

\begin{figure}
 \centering
  \includegraphics[width=\columnwidth]{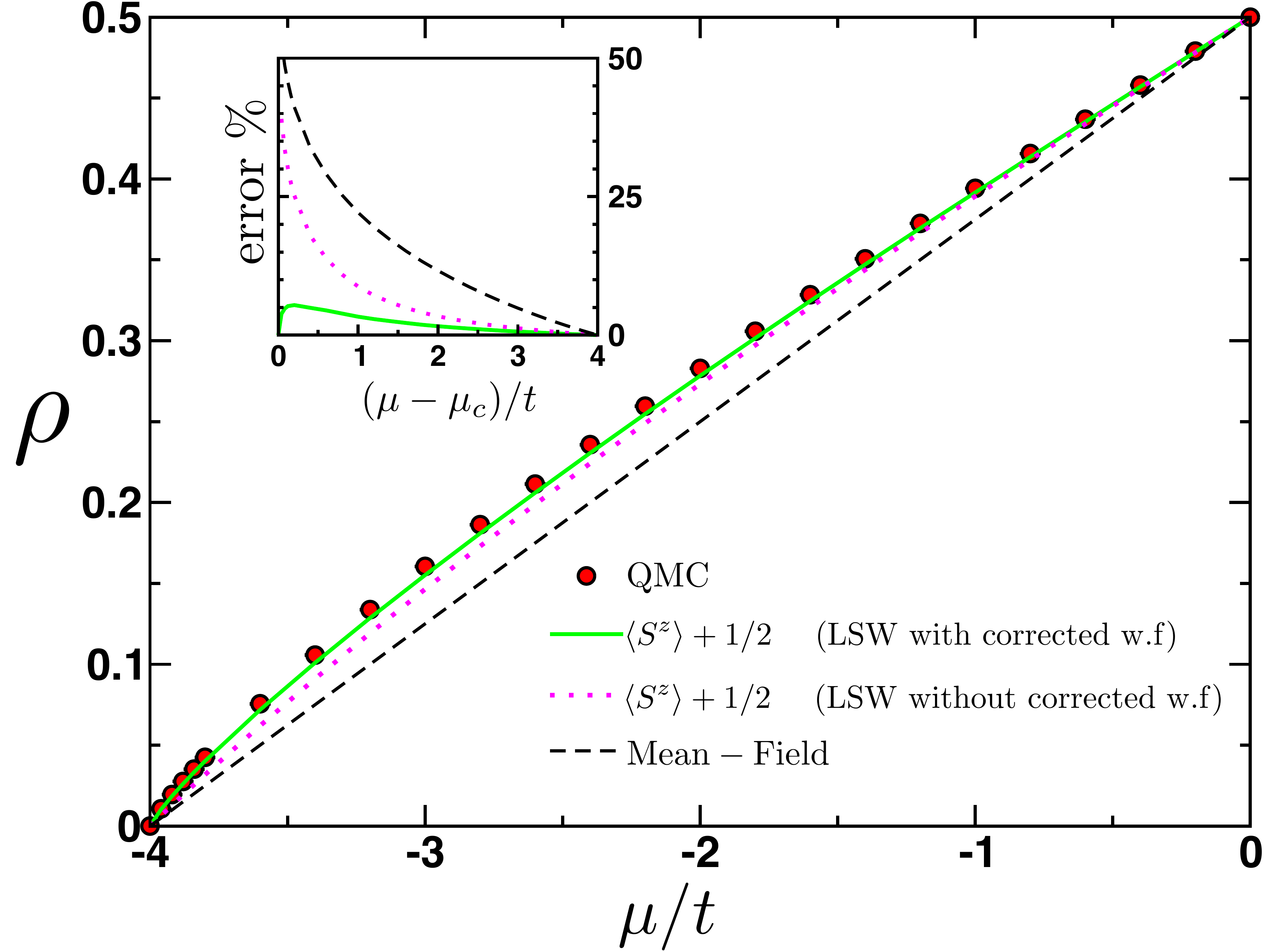}
  \caption{(Color online) Hard-core boson density $\rho$ as a function $\mu/t$. QMC results (symbols) are compared with classical and LSW calculations. The inset shows the relative deviation of the classical and spin-wave results from the numerically exact QMC estimates.}
 \label{fig:RHO}
 \end{figure}
\begin{figure}
 \centering
  \includegraphics[width=\columnwidth]{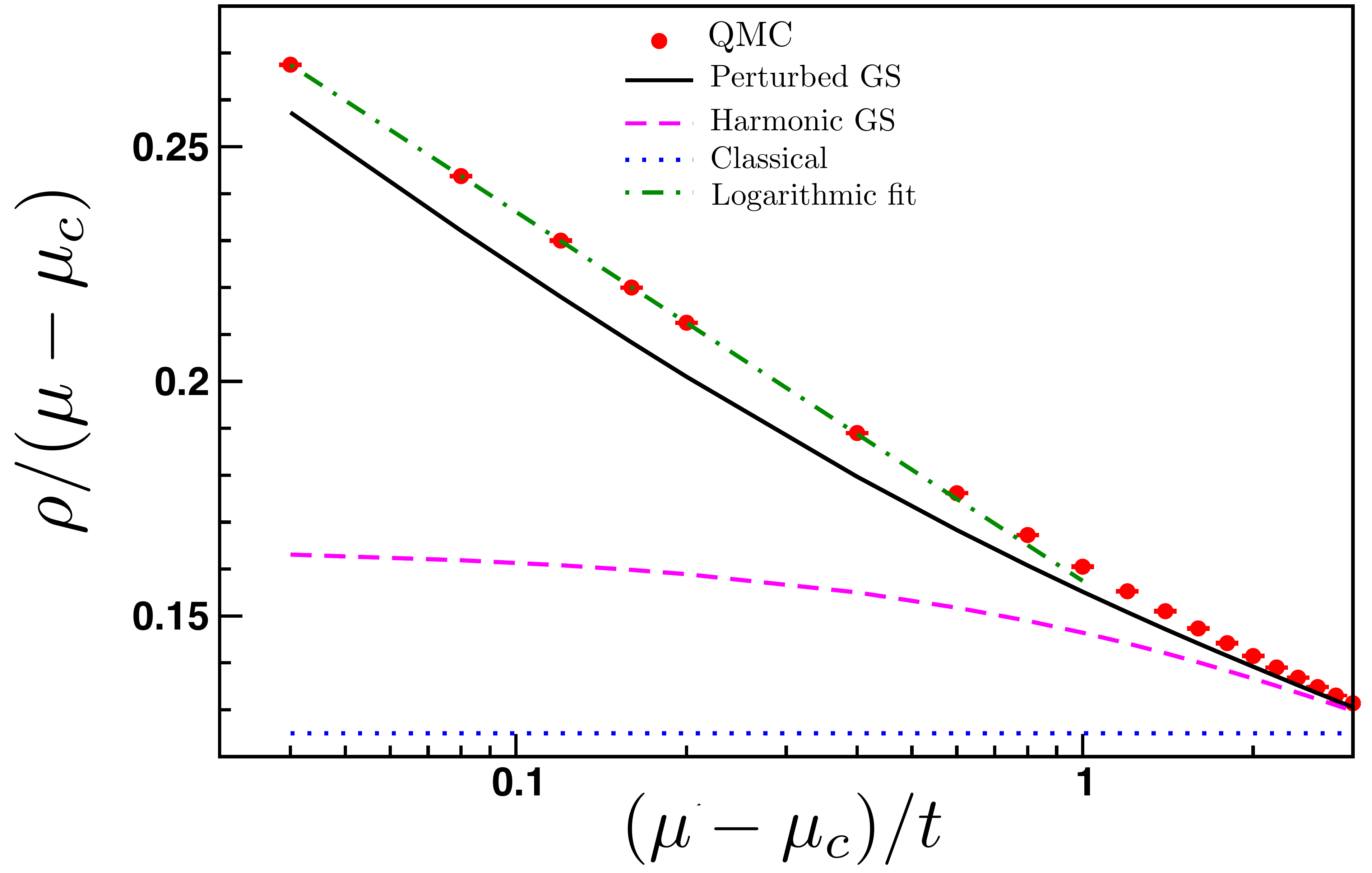}
  \caption{(Color online) Logarithmic corrections to the density as a function of distance from the critical chemical potential $(\mu-\mu_c)$. QMC data (symbols) are described by a fit (dotted-dashed green line) of the form $\alpha\left|\ln\left[\epsilon(\mu-\mu_c)/4t\right]\right|$ with $\alpha\simeq 0.034$ and $\epsilon\simeq 0.04$; the spin-wave calculation using the perturbed ground state (black line) captures the logarithmic correction and yields $\alpha\simeq 0.035$ and $\epsilon\simeq 0.065$. Classical (blue dotted) and LSW results using the non-perturbed harmonic ground-state (magenta dashed) do not capture the logarithmic corrections.}
 \label{fig:RHO_LOG}
 \end{figure}

\subsection{Particle density}
Figure~\ref{fig:RHO} is a plot of the hardcore boson density as a function of $\mu/t$.
The spin wave results for the density are plots of $\langle S^z\rangle+1/2$
calculated using the perturbed ground state (solid line) and using the harmonic ground state (dotted line).
While both approaches yield significant corrections to the mean field density, the expectation value $\langle S^z\rangle$ calculated in the
harmonic ground state misses some terms of order $O(1)$, i.e. $1/S$ corrections to the classical results, as discussed in Sec.~\ref{sec:Averages in the corrected gs}.
This effect is best seen in the inset of Fig.~(\ref{fig:RHO}), which shows the relative deviation of the spin wave results from the QMC estimates.
For small densities the relative deviation from the QMC result is as large as $40\%$ if $\langle S^z\rangle$ is computed using the harmonic ground state.
This deviation never exceeds $5\%$ if the perturbed ground state is used.
Furthermore, the computation in the non-perturbed harmonic ground state misses a very important feature of the dilute Bose gas limit. Indeed, logarithmic corrections have been shown to dominate the low-density limit~\cite{Schick71,Popov72,Fisher88} close to the critical point $\mu_c$, and QMC data are indeed consistent with the behavior $\rho\sim (\mu-\mu_c)\ln{(\mu-\mu_c)}$ with $\mu_c=-4t$, as shown in Figure~\ref{fig:RHO_LOG}. The density computed using the perturbed ground state correctly captures the logarithmic correction whereas $\langle S^z\rangle$ computed in the harmonic ground-state does not.

These results show without any ambiguity that the best way to estimate the density in the context of a semi-classical approximation is to deduce it from the derivative
of the harmonic energy with respect to the chemical potential, or equivalently to deduce it from a calculation of the expectation value of $S^z$ in the ground state that includes leading corrections beyond the harmonic approximation.
The density deduced from the expectation value of $S^z$ calculated in the harmonic ground state is much less accurate,
and qualitatively wrong in the dilute limit.
\subsection{Condensate density and superfluid density}
\subsubsection{LSW and QMC results}
\begin{figure}
 \centering
  \includegraphics[width=\columnwidth]{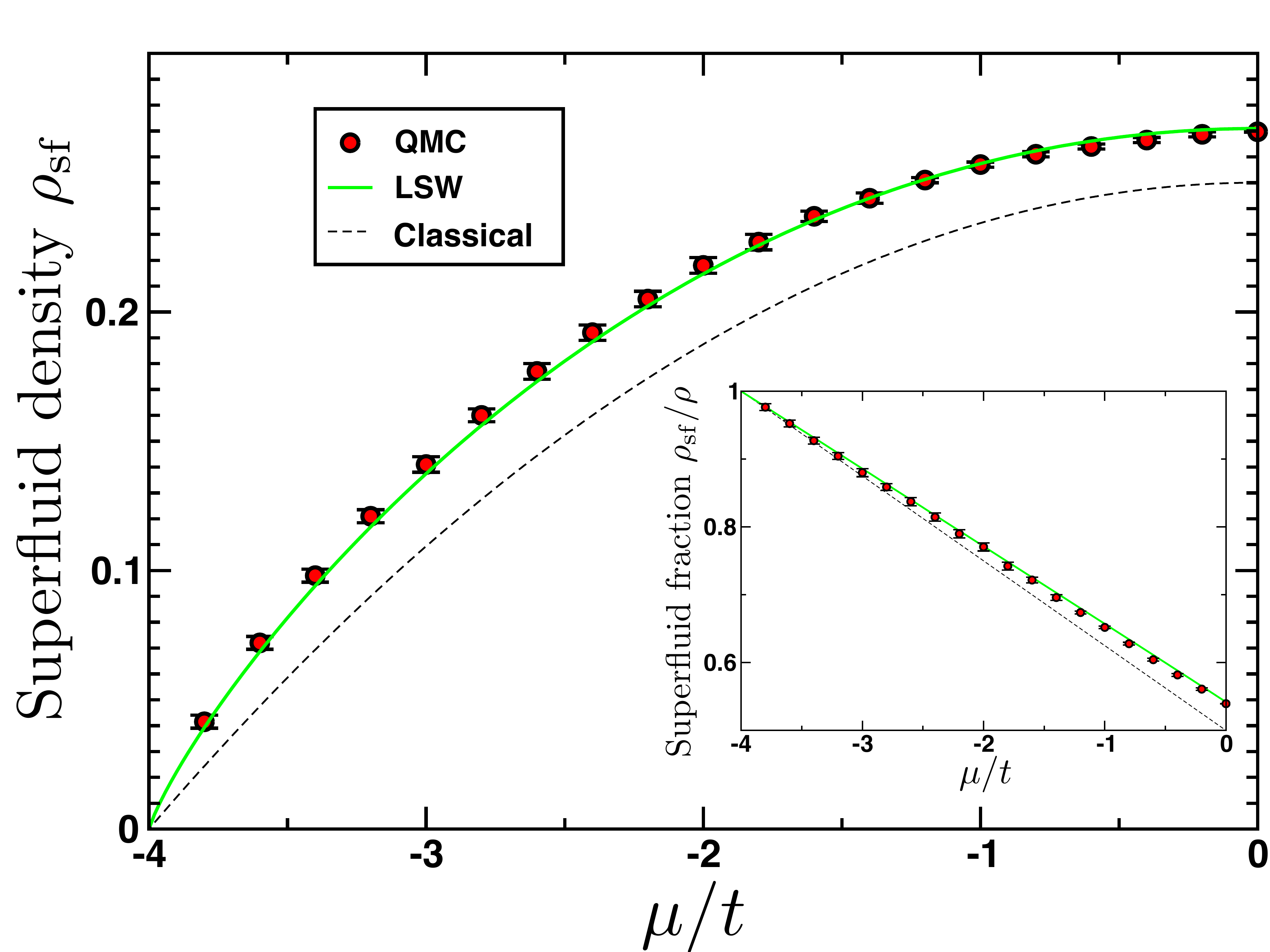}
  \caption{(Color online) Superfluid density as a function of $\mu/t$. Classical (dashed line), semi-classical (solid green line), and QMC (symbols) results are shown. Inset:
superfluid fraction.}
 \label{fig:RHOSF}
 \end{figure}
\begin{figure}
 \centering
  \includegraphics[width=\columnwidth]{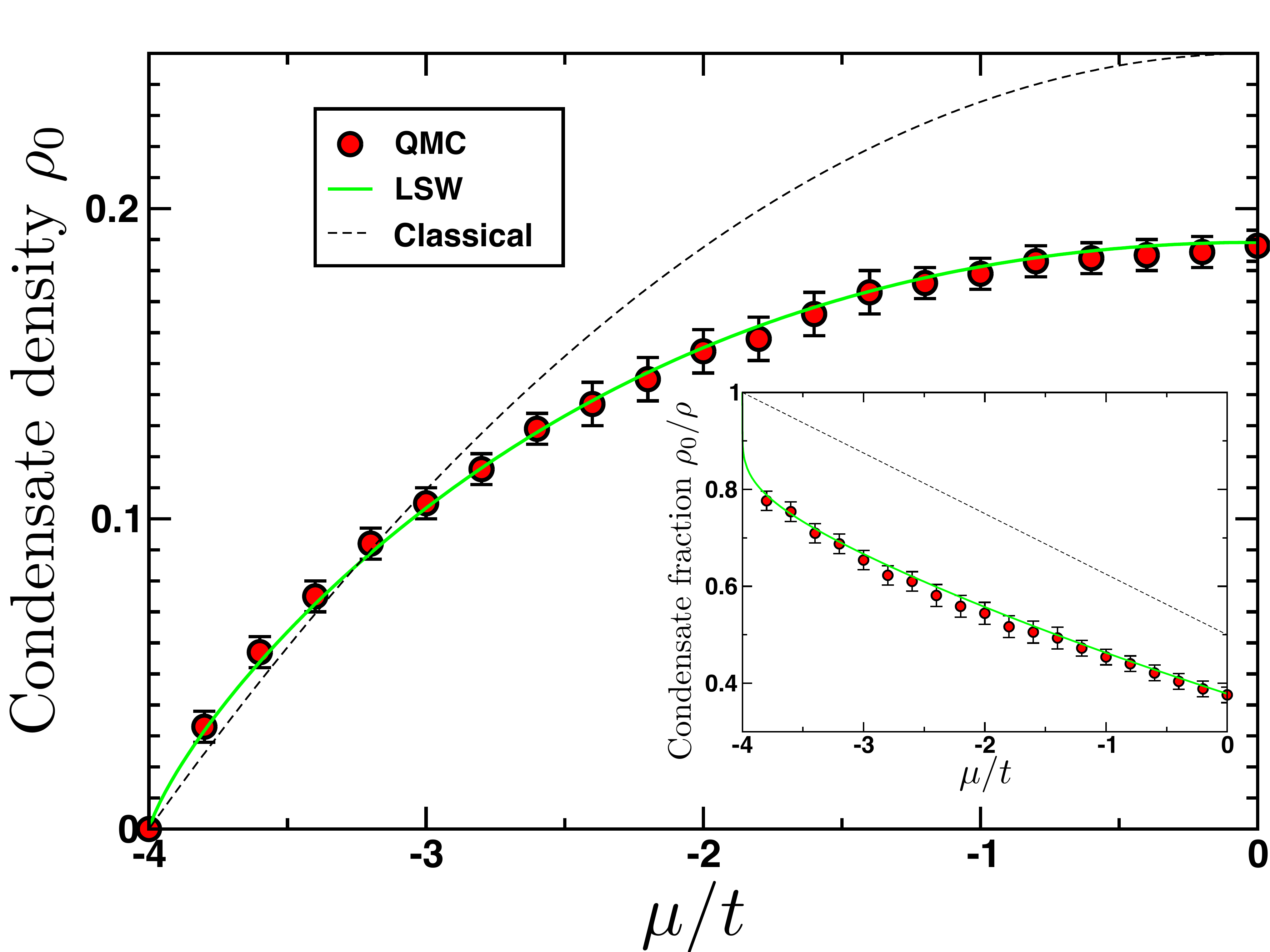}
  \caption{(Color online) Condensate density as a function of $\mu/t$. classical (dashed line), semi-classical (solid green line), and QMC (symbols) results are shown. Inset:
condensate fraction.}
 \label{fig:RHO_0}
 \end{figure}

Figures \ref{fig:RHOSF} and \ref{fig:RHO_0} are plots of the condensate and superfluid densities as a function of $\mu/t$.
The semi-classical results presented in the previous sections are in very good agreement with QMC results, as also discussed in Ref.~\onlinecite{Bernardet02}.
At the classical level, the condensate and the superfluid densities are equal.
The effect of quantum fluctuations is to enhance the superfluidity and to deplete the condensate. QMC estimates for $\rho_{\rm sf}$ and $\rho_0$ are obtained in the directed loop algorithm framework~\cite{Sandvik02} using the winding number fluctuations~\cite{Ceperley87} for $\rho_{\rm sf}$ and the Green's function estimate~\cite{Dorneich01} for $\rho_0$. Note that the latter suffers from larger statistical errors than $\rho_{\rm sf}$.

\subsubsection{Dilute Bose gas limit}

\begin{figure}
 \centering
  \includegraphics[width=\columnwidth]{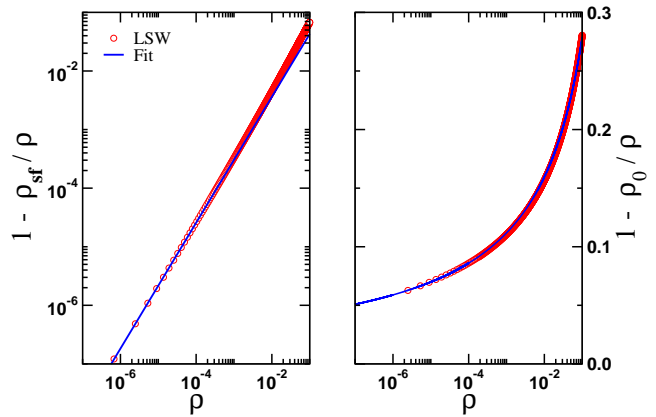}
  \caption{(Color online) Semiclassical results (red symbols) for the superfluid (left) and condensate (right) fractions plotted versus the total density $\rho$. Blue lines are fits of the form Eq.~\eqref{eq:sff} for the superfluid (left) and Eq.~\eqref{eq:cf} for the condensate (right).}
 \label{fig:fractions}
 \end{figure}

Building on the fact that the semiclassical results are very accurate, we analyze the extremely dilute limit for condensed and superfluid fractions using this approximate framework with the help of semi-classical calculations on finite square lattices of linear length $L=10^5$ down to very low particle density of $\rho=10^{-6}$. Such a limit is simply impossible to access using QMC simulations, where the computational cost grows very fast, like $\sim L^4$, so that only systems with a linear size of the order of $L\sim 10^2$ can be accessed.
In both figures \ref{fig:RHOSF} and \ref{fig:RHO_0}, these fractions are shown in the insets. While both fractions are the same at the classical level, the effect of quantum fluctuations is qualitatively different in the two cases. In the extreme dilute limit, they converge to $1$ very differently. Let us first consider the superfluid density. The superfluid fraction $\rho_{\rm sf}/\rho$ is enhanced by quantum fluctuations with respect to the classical case, as seen in the inset of Fig.~\ref{fig:RHOSF}. More precisely, in the dilute limit, the semiclassical superfluid fraction (Fig.~\ref{fig:fractions} left) tends to $1$ like:
\be
\rho_{\rm sf}/\rho = 1- \left(\zeta^2\rho\right)^{\upsilon},
\label{eq:sff}
\ee
with $\zeta\simeq 0.728$, and an exponent $\upsilon\simeq 1.07$ very close to one.
Note that, at the classical level, both fractions (superfluid and condensate) tend to $1$ like $f=1-\rho$.

By contrast to the superfluid fraction, the condensed fraction is more affected by quantum fluctuations in the dilute limit. Indeed, it converges much more slowly to unity, as can be seen in the right panel of Fig.~\ref{fig:fractions} and in the inset of Fig.~\ref{fig:RHO_0}. As first predicted by Schick in Ref.~\onlinecite{Schick71}, logarithmic corrections of the form
\be
\rho_{0}/\rho = 1- \frac{\alpha}{\left|\ln\left(\xi^2\rho\right)\right|}.
\label{eq:cf}
\ee
are expected in the extreme dilute limit.
Fig.~\ref{fig:fractions} (right) shows semi-classical results for the very slow convergence of the condensate fraction to $1$, with a fit to Eq.~\eqref{eq:cf} with $\alpha\simeq 0.86$ and $\xi\simeq 0.68$.
Interestingly we observe that the effective distances $\zeta\sim \xi \sim 0.7$.

%

%
\subsection{Momentum distribution}
\begin{figure}
 \centering
  \includegraphics[width=\columnwidth]{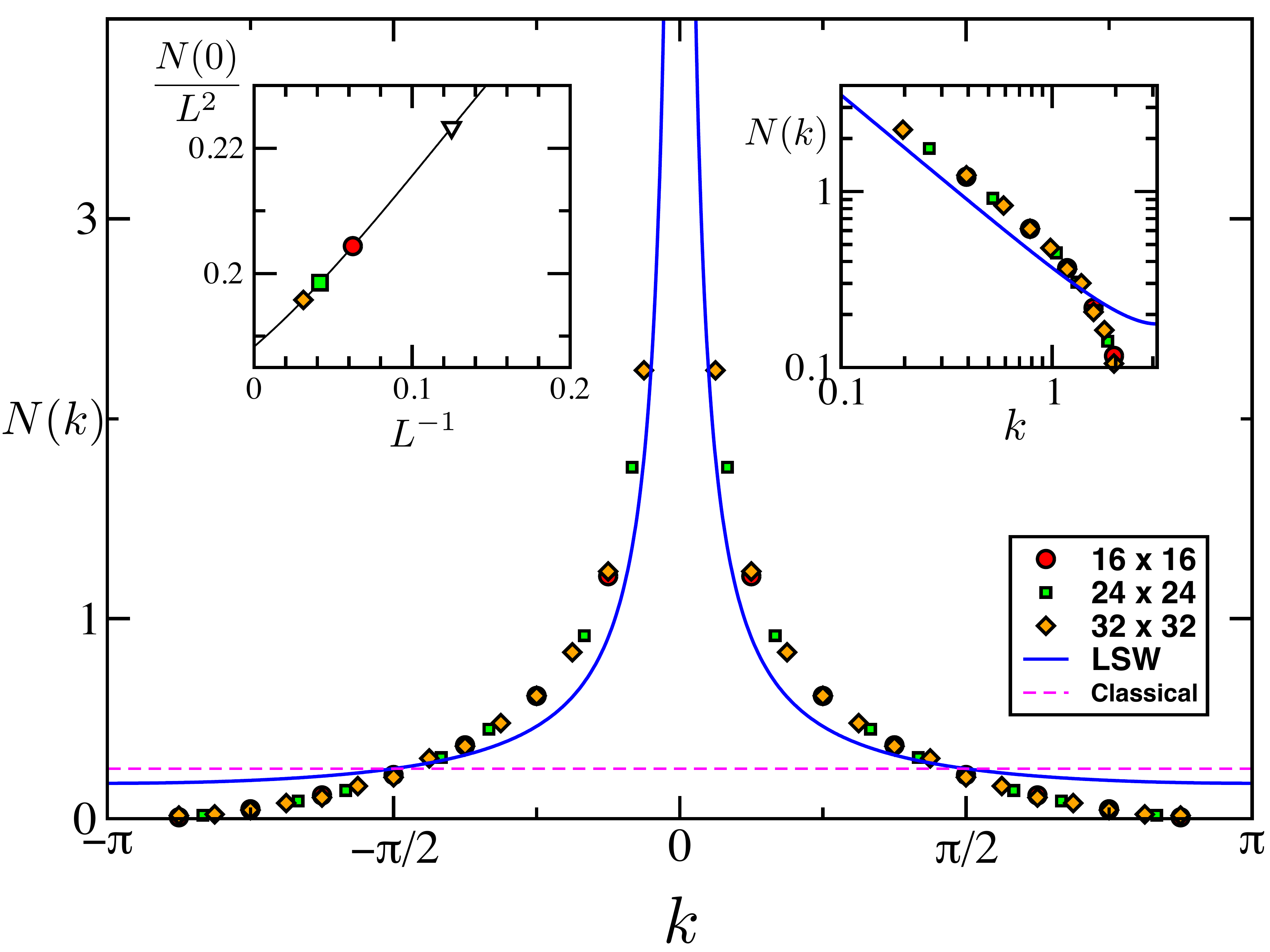}
  \caption{(Color online) Momentum distribution of hard-core bosons $N(\veck)$ Eq~\eqref{eq:NK} at half-filling ($\mu=0$) along the line $k_x=k_y=k$ for $k\neq 0$. Classical (dashed line), semiclassical (full line), and QMC (different symbols for $L=16,~24,~32$) results are shown together.
  Right inset: Log-log plot of the momentum distribution, which diverges as $\sim 1/k$. Left inset: Finite size scaling of the condensate density from QMC as a function of $1/L$. It is well accounted for by a quadratic fit (black line).}
 \label{fig:NK}
 \end{figure}

We now turn to the momentum distribution
\be
N(\veck)=\langle a^{\dagger}_{\veck}a^{\dagga}_{\veck}\rangle,
\label{eq:NK}
\ee
which can be efficiently computed using QMC simulations, following Ref.~\onlinecite{Dorneich01}. Results for the half-filled case ($\mu=0$) are shown in Fig.~\ref{fig:NK} for $k\neq 0$ along the line $k_x=k_y=k$ in the first Brillouin zone.
At the classical level (dashed line), the distribution does not depend on momentum and is equal to $(\rho^\textrm{class.})^2$.
The effect of spin wave fluctuations (solid line) is to introduce a momentum dependence which is singular near
$k=0$ and diverges like $1/k$,
as discussed in section \ref{sec:Averages in the corrected gs}. This redistribution of spectral weight is due to the fact that spin waves deplete the condensate at $k=0$.
The semi-classical calculation of the momentum distribution reproduces the behavior of the QMC results for small $k$.
This is best seen in the right inset of the figure, which is a log-log plot of the distribution showing the
$ 1/k$
dependence of the QMC results for small $k$. Looking back at Eq.~\eqref{eq:NKSW}, the $1/k$ divergence of $N(k)$ is a consequence of the linear spectrum at small momentum $\Omega_{k}\sim k$.
Away from the condensation vector $k=0$, the agreement between semiclassical and QMC results is less good. The QMC estimate is consistent with a distribution that goes to zero when $k\to\pi$ while, according to the semiclassical results, the distribution is only slightly renormalized downwards with respect to the classical constant value.

The contribution at $k=0$ is not shown on the main panel of Fig.~\ref{fig:NK} since it diverges with the system size like $L^2$. In the left inset however we show the QMC result for $\rho_0=N(0)/L^2$ plotted against $1/L$. Using a quadratic fit, we extract the thermodynamic limit value of the condensate density $\rho_0= 0.188(2)$, in good agreement with the estimate $0.191(2)$ reported by Sandvik and Hamer in Ref.~\onlinecite{Sandvik99} (see also Table~\ref{tab:1}).

\section{Density sum rule}\label{sec:Sum rule}

\begin{figure}
 \centering
  \includegraphics[width=\columnwidth]{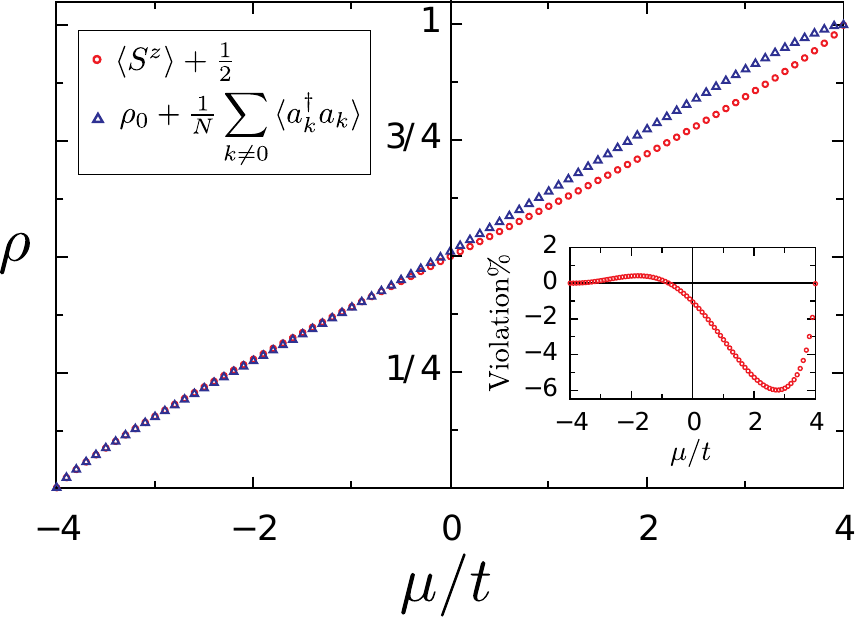}
  \caption{(color online) Total density $\rho$ (red diamonds) and
  $\rho_0+\frac{1}{N}\sum_{\veck\neq 0}{a_\veck^\dagger a_{\veck}}$ (blue triangles) as a function of the chemical potential.
	   The inset presents the violation to the sum rule.}
 \label{fig:Violation sum rule}
 \end{figure}

Having obtained the first order corrections to the total density of particles, the density of condensed particles and the density of uncondensed particles, it is
natural to test the semiclassical approximation with respect to the following sum rule:
\begin{equation}\label{eq:Sum rule}
 \rho=\rho_0+\frac{1}{N}\sum_{\veck\neq0}{a_\veck^\dagger a^\dagga_\veck}.
\end{equation}
The above equality follows directly from the conservation of the number of particles: the total number of hardcore bosons in the system is equal to the sum of the number of condensed and uncondensed particles.
Somewhat surprisingly, with the semiclassical expressions derived in the previous section, the sum rule of Eq.~(\ref{eq:Sum rule}) is violated.
Fig.~\ref{fig:Violation sum rule} shows the average particle density obtained from Eq.~(\ref{eq:Corrected Sz}) and a plot of the sum
$\rho_0+(\sum_{\veck\neq0}a_\veck^\dagger a_\veck^\dagga)/N$ (sum of Eqs. (\ref{eq:Corrected Condensate}) and (\ref{eq:Corrected density of uncondensed particles})).
In the inset, the violation of the sum rule, defined as the relative difference between the two quantities, is shown in the entire filling range. It is smaller than $2\%$ up to half-filling, and never exceeds $6.5\%$ above half-filling.

The origin of the violation of the sum rule is actually quite simple. In the spin-1/2 language, the sum rule relies
on two identities:
\begin{equation}\label{eq:New sum rule}
\sum_i{S_i^+S_i^-}=\frac{1}{N}\sum_{i,j}\sum_{\veck}{S_i^+S_j^-e^{i\veck(\vecr_i-\vecr_j)}}
\end{equation}
and
\be
S_i^z+1/2=S_i^+S_i^-
\ee
which lead to
\begin{equation}\label{eq:New sum rule}
\underbrace{\frac{1}{N}\sum_i n_i}_{\rho}=\underbrace{\frac{1}{N^2}\sum_{i,j}{S_i^+S_j^-}}_{\rho_0}+\underbrace{\displaystyle \frac{1}{N^2}\sum_{\substack{i,j\\\veck\neq0}}{S_i^+S_j^-e^{i\veck(\vecr_i-\vecr_j)}}}_{\frac{1}{N}\sum_{\veck\neq 0}{a_\veck^\dagger a^\dagga_\veck}}
\end{equation}
since, according to the Matsubara-Matsuda transformation, the local density is related to the $z$ component
of the spin by $n_i=S_i^z+1/2$. However, when the spin is larger than 1/2, the identity $S_i^z+1/2=S_i^+S_i^-$
is no longer valid, and the expectation value of $S_i^z$ is no longer equal to that of $S_i^+S_i^--1/2$.

Remarkably enough, in spite of that, the sum rule is almost satisfied, especially below half-filling. This
provides some additional confidence in the accuracy of the semi-classical approach. As a further test, we discuss
in the next section the effect of higher order corrections on the ground state energy.

\section{Second order spin wave theory}
\label{sec:S2}
\subsection{Energy}

The convergence of the expansion of spin operators in terms of Holstein Primakoff bosons Eq.~(\ref{eq:HP transformation}) can be
a cause of concern in the case $S=1/2$.
In this section, we compute the $1/S^2$ correction to the expectation value of several observables.

The ground state energy of the Hamiltonian to $\mathcal{H}^{(2)}$ gives the energy of the original problem to
order $1/S$.
The $1/S^2$ correction comes from both $\mathcal{H}^{(3)}$ and $\mathcal{H}^{(4)}$.
It is obtained by treating $\mathcal{H}^{(3)}$ up to second order in perturbation theory, and $\mathcal{H}^{(4)}$
to first order, as pointed out in another context by Zhitomirsky and Nikuni\cite{Zhitomirsky98}.
The contribution of $\mathcal{H}^{(4)}$ is simply given by $E^{(4)}=\langle\mathcal{H}^{(4)}\rangle/N$,
where the expectation value is calculated in the harmonic ground state. Using Wick's theorem, this contribution
is given by:
\begin{equation}\label{eq:E4}
\begin{array}{lll}
 E^{(4)}&=&\displaystyle\lim_{\Gamma\rightarrow0}\frac{t}{2S^2}\left[\cos^2\theta(2m-\delta)(n-\Delta)\right.\\[3mm]
        & &\quad \left. +(2m+\delta)(\Delta+n)-\sin^2\theta(4m^2+\Delta^2+n^2)\right]
\end{array}
\end{equation}
where $m,n,\delta$ and $\Delta$ are defined by:
\begin{equation}\label{eq:coefficients E4}
 \begin{array}{ll}
  \displaystyle m=\frac{1}{N}\sum_\veck v_\veck^2=\langle b_i^\dagger b_i\rangle &  \displaystyle n=\frac{1}{N}\sum_\veck v_\veck^2\gamma_\veck=2\langle b_i^\dagger b_j\rangle \\
  \displaystyle \delta=\frac{1}{N}\sum_\veck u_\veck v_\veck=-\langle b_i b_i\rangle& \displaystyle \Delta=\frac{1}{N}\sum_\veck u_\veck v_\veck\gamma_\veck=-2\langle b_i b_j\rangle
 \end{array}
\end{equation}
where $i$ and $j$ are nearest neighbors.

Being odd in the number of bosonic operators, $\mathcal{H}^{(3)}$ contributes only at second order in non degenerate perturbation theory:
\begin{equation}
 E^{(3)}=\frac{1}{N}\sum_{|e\rangle\neq|0\rangle}{\frac{|\langle e|\mathcal{H}^{(3)}|0\rangle|^2}{E_{|0\rangle}-E_{|e\rangle}}}\sim\mathcal{O}\left(\frac{1}{S^2}\right)
\end{equation}
where $E^{(3)}$ is the energy per site.
The effect of $\mathcal{H}^{(3)}$ on the Bogoliubov vacuum is to create either
single magnon excited states or three magnon excited states. We treat these two cases independently
and write the $\mathcal{H}^{(3)}$ contribution to the energy as $E^{(3)}=E^{(3)}_{\textrm{1m}}+E^{(3)}_{\textrm{3m}}$.
The single magnon component of $\mathcal{H}^{(3)}|0\rangle$ is:
\begin{equation}
\frac{2t \sin\theta\cos\theta}{S\sqrt{2S}\sqrt{N}} \sum_\veck(u_0-v_0)\left[2v_\veck^2+\gamma_\veck(v_\veck^2-u_\veck v_\veck)\right]|1_{\vecq=0}\rangle
\end{equation}
which leads to:
\begin{equation}
\begin{array}{lll}
  E^{(3)}_{\textrm{1m}}&=&\displaystyle\lim_{\Gamma\rightarrow0}{\frac{-t^2}{S^2}\sin^2\theta\cos^2\theta\frac{(u_0-v_0)^2}{\sqrt{A_0^2-B_0^2}}} \cdot\\
	    & &\displaystyle\quad\frac{1}{N^2}\left|\sum_\veck\left[2v_\veck^2+\gamma_\veck(v_\veck^2-u_\veck v_\veck)\right]\right|^2 \\
                           &=&\displaystyle-\frac{t}{2S^2}\cos^2\theta_0\lim_{\Gamma\rightarrow0}(2m+n-\Delta)^2
\end{array}
\end{equation}

In the thermodynamic limit $E^{(3)}_\textrm{3m}$ is dominated by the excited states in which the three
magnons all have different momenta.
The three-magnon component of $\mathcal{H}^{(3)}|0\rangle$ that fulfills this condition is given by:
\begin{equation}
\begin{array}{c}
\displaystyle \frac{2t \sin\theta\cos\theta}{S\sqrt{2S}\sqrt{N}}\sum_{\veck,\vecq}{^\prime\textrm{f}(\vecq,\veck)|1_\veck1_\vecq1_{-\veck-\vecq}\rangle} \\
\\
\displaystyle \textrm{f}(\vecq,\veck)=\gamma_\vecq(u_{\veck+\vecq}v_\veck v_\vecq-u_\veck u_\vecq v_{\veck+\vecq})
\end{array}
\end{equation}
where the sum $\sum_{\veck,\vecq}^\prime$ is such that the three momenta $\veck,\vecq$ and $-\veck-\vecq$ are all different.
The three-magnon contribution to $E^{(3)}$ takes the form:
\begin{equation}
\begin{array}{lll}
E^{(3)}_\textrm{3m}&=&\displaystyle\frac{-2t^2}{S^3}\frac{1}{N^2}\sin^2\theta_0\cos^2\theta_0\\[2mm]
                       & &\displaystyle\quad \cdot\lim_{\Gamma\rightarrow0}{\frac{1}{3!}\sum_{\veck,\vecq}\frac{\textrm{F}^2(\veck,\vecq)}{\Omega_\vec\veck+\Omega_\vecq+\Omega_{\veck+\vecq}}}
\end{array}
\end{equation}
where $\textrm{F}(\veck,\vecq)$ is defined by:
\begin{equation}
\begin{array}{lll}
 \textrm{F}(\veck,\vecq)&=&\displaystyle\textrm{f}(\veck,\vecq)+\textrm{f}(-\veck-\vecq,\vecq)+\textrm{f}(\vecq,\veck)+\textrm{f}(\vecq,-\veck-\vecq)\\ [2mm]
                & &\displaystyle +\textrm{f}(-\veck-\vecq,\veck)+\textrm{f}(\veck,-\veck-\vecq).
\end{array}
\end{equation}

\begin{figure}
 \centering
  \includegraphics[width=\columnwidth]{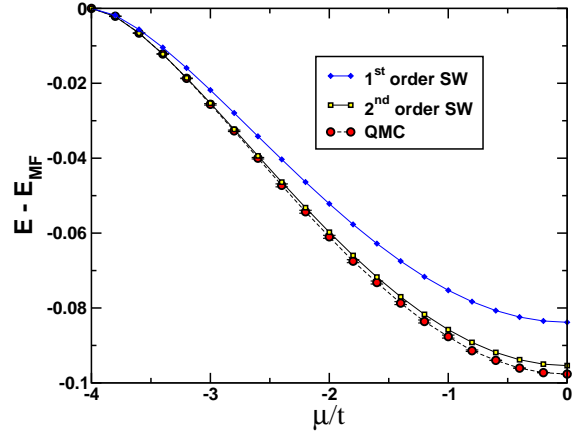}
  \caption{(Color online) Plot of the $1^\textrm{st}$ order spin wave energy, of the $2^\textrm{nd}$
  order spin wave energy and of the QMC energy measured with respect to the mean field energy.}
 \label{fig:E}
 \end{figure}

The spin-wave approximation of the energy per site to order $\mathcal{O}(1/S^2)$
\begin{equation}
 E=\mathcal{E}+E^{(2)}+E^{(3)}_\textrm{1m}+E^{(3)}_\textrm{3m}+E^{(4)}
\end{equation}
is plotted in Fig.(\ref{fig:E}), together with the $1^\textrm{st}$ order SWT and QMC results, as a function of $\mu/t$.
All energies are measured with respect to the classical energy $\mathcal{E}$.
The difference with the classical energy is monotonously increasing (in absolute value) from the dilute limit up to half-filling, where the correction due to quantum fluctuations is the most important. Clearly, the $1/S$ correction captures most the quantum correction, but the inclusion of $1/S^2$ corrections leads to a significantly better
agreement with QMC results. This systematic improvement upon including higher order $1/S$ corrections gives additional support to the semi-classical expansion.

\subsection{Other observables}
Superfluid and condensate densities, as well as the compressibility $\kappa=\partial \rho/\partial\mu$ can also be calculated at order $1/S^2$. The calculation is straightforward but cumbersome, and for simplicity it is not reproduced here. The estimates that we have obtained at half-filling are listed in Table~\ref{tab:1}, together with QMC results from the present work as well as from Ref.~\onlinecite{Sandvik99}. For the ground-state energy and the compressibility, the agreement increases systematically from $1/S^0$ (classical) to $1/S^2$. For the energy, the relative error is $\sim 9\%$ for the classical estimate, less than $1.3\%$ including the $1/S$ correction, and $\sim 0.2\%$ including the $1/S^2$ correction. For the condensate density, the result including the $1/S$ correction is already within the error
bars of QMC, and it is not clear whether including $1/S^2$ correction leads to any improvement. For the superfluid
density, both the results including corrections up to order $1/S$ and $1/S^2$ lie outside the error bars of QMC, and
the result up to order $1/S$ appears to be better than the result up to order $1/S^2$. In any case, the improvement over the classical result is clear for all quantities.

\begin{table}[!ht]
\begin{tabular}{c|l|l|l|l}
&$E_{\rm gs}$& $\rho_{\rm sf}$ & $\rho_0$ & $\kappa$\\
\hline
\hline
Classical&-1&0.25&0.25&0.125\\
\hline
$1/S$ SW&-1.08382&0.27095&0.18904&0.1075\\
\hline
$1/S^2$ SW&-1.09539&0.27198&0.19127&0.1053\\
\hline
QMC (\scriptsize Ref.~\onlinecite{Sandvik99})&-1.097648(4)&0.2696(2)&0.191(2)&0.1048(1)\\
\hline
QMC {\scriptsize{(this work)}}&-1.09764(1)&0.2697(2)&0.188(2)&0.1048(1)\\
\hline
\end{tabular}
\caption{\label{tab:1}Ground-state estimates at half-filling ($\mu=0$) for the energy per site $e_0$, the superfluid density $\rho_{\rm sf}$, the condensate density $\rho_0$, and the compressibility $\kappa$.
The three first lines are analytical results from classical and spin-waves at first ($1/S$ SW) and second order ($1/S^2$ SW).
Below are shown QMC estimates from SSE simulations obtained by Sandvik in Ref.~\onlinecite{Sandvik99} and in this work.}
\end{table}

\section{Conclusion}\label{sec:Conclusion}

The semi-classical approach to hard-core bosons on a lattice, which is based on a large $S$ approximation
to the Matsubara-Matsuda spin-1/2 version of the Hamiltonian, has been revisited, with a few questions in mind:
What is the correct way to get the exact $1/S$ correction to various observables? Can the method be extended to
a more complete characterization of ground state correlations? How good is the semi-classical approach in dealing with
some of the subtleties of bosons in 2D, for instance the logarithmic corrections of the dilute limit? We have
shown that to get the exact $1/S$ correction to the ground state expectation value of various observables, it
is necessary to include corrections to the harmonic ground state, and we have explicitly shown how to include them
for the density, the condensate, and the momentum distribution function, for which, to the best of our knowledge,
we have provided the first semi-classical expression. By a careful comparison with QMC  results, we have shown that, when it is done properly, the semi-classical expansion is remarkably accurate.
In particular, we have shown that it reproduces the logarithmic corrections predicted a long time ago in the dilute limit, as well as the divergence of the momentum distribution at $k=0$. We have further tested the reliability of the $1/S$ results by looking at the density sum rule and at higher order corrections. Whichever way one looks at it,
the semi-classical approach appears as a very accurate description of hard-core bosons on a lattice.
\section*{Acknowledgments}
We are grateful to George Batrouni for very useful discussions about
the results of Ref.~\onlinecite{Bernardet02}. This project has been supported by the Swiss National Fund
and by MaNEP.

\newpage
\appendix
\section{Computation of the superfluid density}
\label{sec:app}
\subsection{Classical value}
The superfluid density can be obtained by imposing a phase gradient
$\Phi_{{{\vecr}_i}+{\bf e}}-\Phi_{{\vecr}_i}=\varphi$ to the system (${\bf e}$ being the unit vector along the $x$ or $y$ axis of the lattice).
At the classical level, this leads to the following energy cost per site
\be
E(\varphi)-E(0)=t\left({\sin\theta_0}\right)^2\varphi^2+O(\varphi^4).
\ee
Using the analogy introduced by Fisher, Barber and Jastrow in Ref.~\onlinecite{Fisher73} where the kinetic energy density of a superflow
(density $\rho_{\rm sf}$ and velocity $v_{\rm sf}$) in one direction is $\delta E({v_{\rm sf}})=\frac{1}{2}m^*\rho_{\rm sf}\left({v_{\rm sf}}\right)^2$,
with $v_{\rm sf}=(\hbar/m^*)\varphi$. This gives for the superfluid density
\be
\rho_{\rm sf}=\frac{m^*}{2\hbar^2}\Upsilon_{\rm sf},
\ee
where $\Upsilon_{\rm sf}=\partial^2 E(\varphi)/\partial \varphi^2\bigr|_{\varphi=0}$ is the helicity modulus, the effective mass is given by
$2m^*/\hbar^2=1/(2t)$, and the factor of 2 comes from the fact that the twist $\varphi$ has been introduced in both directions. Finally we get at the classical level
\be
\rho_{\rm sf}=\frac{\sin^2\theta_0}{4}.
\ee
Interestingly, we remark that condensate and superfluid densities are equal at this level of approximation
\subsection{SW corrections}
In order to evaluate the SW corrections to the superfluid fraction,
the phase gradient can be introduced directly on the bosonic operators
\bea
a^{\dagger}_{\vecr_i}&\to&a^{\dagger}_{\vecr_i}~{\rm{e}}^{i\Phi_{\vecr_i}}\nonumber\\
a_{\vecr_i}&\to&a_{\vecr_i}^{\vphantom{\dagger}}~{\rm{e}}^{-i\Phi_{\vecr_i}},
\eea
which in term of equivalent spin operators translates into
\bea
S_{\vecr_i}^{x}\to S_{\vecr_i}^{x}\cos \Phi_{\vecr_i}-S_{\vecr_i}^{y}\sin \Phi_{\vecr_i}\nonumber\\
S_{\vecr_i}^{y}\to S_{\vecr_i}^{x}\sin \Phi_{\vecr_i}+S_{\vecr_i}^{y}\cos \Phi_{\vecr_i}.
\eea
Therefore, the rotation \eqref{eq:Local Spin Rotation} becomes
\bea
S^x_{\vecr_i}&=&\left(\cos\theta S^u_{\vecr_i} +\sin\theta S^w_{\vecr_i}\right)\cos \Phi_{\vecr_i}-S^v_{\vecr_i}\sin \Phi_{\vecr_i}\nonumber\\
S^y_{\vecr_i}&=&\left(\cos\theta S^u_{\vecr_i} +\sin\theta S^w_{\vecr_i}\right)\sin \Phi_{\vecr_i}+S^v_{\vecr_i}\cos \Phi_{\vecr_i}\nonumber\\
S^z_{\vecr_i}&=&-\sin\theta S^u_{\vecr_i} +\cos\theta S^w_{\vecr_i}.
\eea
In the new rotated frame, at the linear SW approximation the XY Hamiltonian now reads
%
\bea
{\cal{H}}^{(2)}(\varphi)&=&2\sum_{\veck}\Bigl[A_{\veck}(\varphi)\left(a_{\veck}^{\dagger}a_{\veck}+a_{-\veck}^{\dagger}a_{-\veck}\right)\nonumber\\
&+&{{B}}_{\veck}(\varphi)\left(a_{\veck}^{\dagger}a^{\dagger}_{-\veck}+a_{\veck}a_{-\veck}\right)\Bigr],
\eea
with
\be
A_{\veck}(\varphi)=-\frac{t}{2}\cos\varphi\left[(1+\cos^2\theta)\gamma_{\veck}-4\right],\ee
and
\be{{B}}_{\veck}(\varphi)=\frac{t}{2}\cos\varphi(\sin^2\theta)\gamma_{\veck},\ee
where we used S=1/2 and where $\theta$ is fixed by the equation $\cos\theta=\mu/(4t\cos\phi)$ imposed by the minimization of the classical energy.
At order $\varphi^2$ we have of course $\cos\varphi\simeq 1-\varphi^2/2$. Similarly the condition on $\theta$ yields
$\cos^2\theta\simeq\cos^2\theta_0\left(1+\varphi^2\right)$, and $\sin^2\theta\simeq \sin^2\theta_0-\cos^2\theta_0\varphi^2$.
Therefore we have up to the order $\varphi^2$:
\be
A_{\veck}(\varphi)=A_{\veck}(0)-\frac{\varphi^2}{2}\left(A_{\veck}(0)+{t\gamma_\veck\cos^2\theta_0}\right)
\ee
and
\be
{{B}}_{\veck}(\varphi)={{B}}_{\veck}(0)-\frac{\varphi^2}{2}\left({{B}}_{\veck}(0)+{t\gamma_\veck\cos^2\theta_0}\right).
\ee
Writing $A_{\veck}(0)=A_{\veck}$ and $B_{\veck}(0)=B_{\veck}$, the $1/S$ correction to the GS energy in the presence of a small twist reads
\bea
E^{(2)}(\varphi)-E^{(2)}(0)&=&\frac{\varphi^2}{N}\sum_{\veck}\Bigl\{2t-\sqrt{A_{\veck}^2-{{B}}_{\veck}^2}\nonumber\\
&-&{t\gamma_\veck\cos^2\theta_0}\sqrt{\frac{A_{\veck}-{{B}}_{\veck}}{A_{\veck}+{{B}}_{\veck}}}\Bigr\}.\nonumber
\eea
\\
Thus the superfluid density is given at $1/S$ order by the following expression
\bea
\rho_{\rm sf}&=&\frac{\sin^2\theta_0}{4}\nonumber\\
&+&\frac{1}{4Nt}\sum_{\veck}\Bigl\{2t-\sqrt{A_{\veck}^2-{{B}}_{\veck}^2}\nonumber\\
&-&{t\gamma_\veck\cos^2\theta_0}\sqrt{\frac{A_{\veck}-{{B}}_{\veck}}{A_{\veck}+{{B}}_{\veck}}}\Bigr\}.
\eea

This expression and that of Ref.~\onlinecite{Bernardet02} are strictly equivalent
only at $\mu=-4t$ (low density limit) and at $\mu=0$ (half filling).
In the range $-4<\mu/t<0$ and $0<\mu/t<4$ they differ by a term which is of order $1/S^2$ \footnotemark[\value{footnote}].
\section{momentum distribution}
\label{sec:app2}
We start by expressing $S_i^+S_j^-$ in the rotated frame:
\begin{equation}
\begin{array}{lll}
 S_i^+S_j^-&=& \cos^2\theta S_i^{x^\prime}S_j^{x^\prime}+\sin^2\theta S_i^{z^\prime}S_j^{z^\prime} +S_i^{y^\prime}S_j^{y^\prime} \\
           & &+\cos\theta\sin\theta\left(S_i^{x^\prime}S_j^{z^\prime}+S_i^{z^\prime}S_j^{x^\prime}\right)\\
           & &+\cos\theta S_i^{z^\prime}\delta_{i,j}-\sin\theta S_i^{x^\prime}\delta_{i,j}.
\end{array}
\end{equation}\\
The terms involved in $\langle S_i^+S_j^-\rangle$ up to order $\mathcal{O}(S)$ are:
\begin{equation}\label{eq:average S+iS-j}
\begin{array}{lll}
 \langle S_i^+S_j^-\rangle &=& \displaystyle \frac{S}{2} \cos^2\theta \langle0|(b_i+b_i^\dagger)(b_j+b_j^\dagger)|0\rangle \\[2mm]
                           & & \displaystyle + S^2\sin^2\theta -S\sin^2\theta \langle0|b_i^\dagger b_i+b_j^\dagger b_j|0\rangle \\[2mm]
                           & & \displaystyle -\frac{S}{2}\langle0|(b_i-b_i^\dagger)(b_j-b_j^\dagger)|0\rangle \\[2mm]
                           & & \displaystyle +\cos\theta\sin\theta\frac{S}{\sqrt{2}}\left(\langle0|(b_i+b_i^\dagger)|\phi^\frac{1}{2}\rangle+\textrm{h.c.}\right)\\[2mm]
                           & & \displaystyle +\cos\theta\sin\theta\frac{S}{\sqrt{2}}\left(\langle0|(b_j+b_j^\dagger)|\phi^\frac{1}{2}\rangle+\textrm{h.c.}\right)\\[2mm]
                           & & \displaystyle +S\cos\theta \delta_{i,j}.
\end{array}
\end{equation}
Injecting this result in Eq.~(\ref{eq:distribution n(k)}) and making use of the definitions of the inverse Fourier transforms of the $b_i$ operators, Eq.~(\ref{eq:Definition FT}),
we obtain:
\begin{equation}\label{eq:FT distribution n(k)}
\begin{array}{ll}
\langle a_\veck^\dagger a_\veck\rangle&=\displaystyle\lim_{\substack{\Gamma\rightarrow0 \\ S \rightarrow \frac{1}{2}}}\Big\{\frac{S}{2}\cos^2\theta \langle0|(b_{-\veck}+b_{\veck}^\dagger)(b_{\veck}+b_{-\veck}^\dagger)|0\rangle \\[2mm]
                    &        \displaystyle + S^2\sin^2\theta N\delta_{\veck,0}\\[2mm]
                    &        \displaystyle -S\sin^2\theta \delta_{\veck,0} \langle0|\sum_{i}{b_i^\dagger b_i (e^{i \veck R_i}+e^{-i \veck R_i})}|0\rangle \\[2mm]
                    &        \displaystyle -\frac{S}{2}\langle0|(b_{-\veck}-b_{\veck}^\dagger)(b_{\veck}-b_{-\veck}^\dagger)|0\rangle \\[2mm]
                    &        \displaystyle +\cos\theta\sin\theta\frac{\sqrt{N}S}{\sqrt{2}}\delta_{\veck,0}\left(\langle0|(b_{-\veck}+b_{\veck}^\dagger)|\phi^\frac{1}{2}\rangle+\textrm{h.c.}\right)\\[2mm]
                    &        \displaystyle +\cos\theta\sin\theta\frac{\sqrt{N}S}{\sqrt{2}}\delta_{\veck,0}\left(\langle0|(b_\veck+b_{-\veck}^\dagger)|\phi^\frac{1}{2}\rangle+\textrm{h.c.}\right)\\[2mm]
                    &        \displaystyle +S\cos\theta \Big\}.
\end{array}
\end{equation}
Hence, the number of particles at momentum $\vec{\veck}\neq0$ is given by
\begin{equation}\label{eq:n(k)}
\begin{array}{lcl}
\langle a_\veck^\dagger a_\veck\rangle&=&\displaystyle\lim_{\substack{\Gamma\rightarrow0 \\ S \rightarrow \frac{1}{2}}}\Big\{\displaystyle  \frac{S}{2}\cos^2\theta \langle0|(b_{-\veck}+b_{\veck}^\dagger)(b_{\veck}+b_{-\veck}^\dagger)|0\rangle \\[2mm]
                              & &\quad\displaystyle -\frac{S}{2}\langle0|(b_{-\veck}-b_{\veck}^\dagger)(b_{\veck}-b_{-\veck}^\dagger)|0\rangle  +S\cos\theta \Big\}\\[3mm]
                              &=&\displaystyle \frac{1}{4}(1+\cos\theta_0)^2\\
                              &+&\left[(1+\cos^2\theta_0)v_\veck^2+u_\veck v_\veck\sin^2\theta_0\right]/2.
                              \end{array}
\end{equation}
\end{document}